\documentclass[acmsmall]{acmart}

\usepackage{graphicx}
\usepackage[utf8]{inputenc}
\usepackage{tabularx}
\usepackage{multirow}
\usepackage{booktabs}
\usepackage{rotating}
\usepackage{CJK}
\usepackage[utf8]{inputenc} 
\usepackage[T1]{fontenc}    
\usepackage{natbib}

\AtBeginDocument{%
  \providecommand\BibTeX{{%
    \normalfont B\kern-0.5em{\scshape i\kern-0.25em b}\kern-0.8em\TeX}}}

\setcopyright{acmlicensed}
\acmPrice{15.00}
\acmDOI{10.1145/3611049}
\acmYear{2023}
\copyrightyear{2023}
\acmSubmissionID{play23main-p3384-p}
\acmJournal{PACMHCI}
\acmVolume{7}
\acmNumber{CHI PLAY}
\acmArticle{403}
\acmMonth{11}
\received{2023-02-21}
\received[accepted]{2023-07-07}

\begin{document}

\title{Fused Spectatorship: Designing Bodily Experiences Where Spectators Become Players}

\author{Rakesh Patibanda}
\email{rakesh@exertiongameslab.org}
\orcid{0000-0002-2501-9969}
\affiliation{%
  \institution{Monash University}
  \department {Exertion Games Lab, Department of Human-Centred Computing}
  \city{Melbourne}
  \country{Australia}
  \postcode{3166}
}

\author{Aryan Saini}
\email{aryan@exertiongameslab.org}
\orcid{0000-0002-2844-3343}
\affiliation{%
  \institution{Monash University}
  \department {Exertion Games Lab, Department of Human-Centred Computing}
  \city{Melbourne}
  \country{Australia}
  \postcode{3166}
}

\author{Nathalie Overdevest}
\email{nathalie@exertiongameslab.org}
\orcid{0000-0002-2022-5480}
\affiliation{%
  \institution{Monash University}
  \department {Exertion Games Lab, Department of Human-Centred Computing}
  \city{Melbourne}
  \country{Australia}
  \postcode{3166}
}

\author{Maria F. Montoya}
\email{maria@exertiongameslab.org}
\orcid{0000-0001-8587-2358}
\affiliation{%
  \institution{Monash University}
  \department {Exertion Games Lab, Department of Human-Centred Computing}
  \city{Melbourne}
  \country{Australia}
  \postcode{3166}
}

\author{Xiang Li}
\orcid{0000-0001-5529-071X}
\email{xl529@cam.ac.uk}
\affiliation{%
  \institution{University of Cambridge}
  \department{Department of Engineering}
  \city{Cambridge}
  \country{United Kingdom}
  \postcode{CB2 1PZ}
}

\author{Yuzheng Chen}
\orcid{0000-0001-9538-5369}
\email{yuzheng.chen18@student.xjtlu.edu.cn}
\affiliation{%
  \institution{Xi'an Jiaotong-Liverpool University}
  \city{Suzhou}
  \country{China}
}

\author{Shreyas Nisal}
\email{shreyasnisal@gmail.com}
\orcid{0000-0002-5066-6885}
\affiliation{%
  \institution{Monash University}
  \department {Exertion Games Lab, Department of Human-Centred Computing}
  \city{Melbourne}
  \country{Australia}
  \postcode{3166}
}

\author{Josh Andres}
\email{josh.andres@anu.edu.au}
\orcid{0000-0001-5882-3139}
\affiliation{%
  \institution{Australian National University}
  \department{School of Cybernetics}
  \city{Acton}
  \country{Australia}
  \postcode{2601}
}

\author{Jarrod Knibbe}
\email{jarrod.knibbe@unimelb.edu.au}
\orcid{0000-0002-8844-8576}
\affiliation{%
  \institution{University of Melbourne}
  \department{School of Computing and Information Systems}
  \city{Parkville}
  \country{Australia}
  \postcode{3010}
}

\author{Elise van den Hoven}
\email{Elise.VandenHoven@uts.edu.au}
\orcid{0000-0002-0888-1426}
\affiliation{%
  \institution{University of Technology Sydney}
  \city{Sydney}
  \country{Australia}}
\affiliation{%
  \institution{Eindhoven University of Technology}
  \city{Eindhoven}
  \country{Netherlands}}

\author{Florian `Floyd' Mueller}
\email{floyd@exertiongameslab.org}
\orcid{0000-0001-6472-3476}
\affiliation{%
  \institution{Monash University}
  \department {Exertion Games Lab, Department of Human-Centred Computing}
  \city{Melbourne}
  \country{Australia}
  \postcode{3166}
}
\renewcommand{\shortauthors}{Patibanda, Saini, Overdevest, F. Montoya, Li, Chen, Nisal, Andres, Knibbe, van den Hoven and Mueller.}

\begin{abstract}
  Spectating digital games can be exciting. However, due to its vicarious nature, spectators often wish to engage in the gameplay beyond just watching and cheering. To blur the boundaries between spectators and players, we propose a novel approach called ``Fused Spectatorship'', where spectators watch their hands play games by loaning bodily control to a computational Electrical Muscle Stimulation (EMS) system. To showcase this concept, we designed three games where spectators loan control over both their hands to the EMS system and watch them play these competitive and collaborative games. A study with 12 participants suggested that participants could not distinguish if they were watching their hands play, or if they were playing the games themselves. We used our results to articulate four spectator experience themes and four fused spectator types, the behaviours they elicited and offer one design consideration to support each of these behaviours. We also discuss the ethical design considerations of our approach to help game designers create future fused spectatorship experiences.
\end{abstract}




\maketitle

\section{Introduction}\label{sec:Intro}
Parallel to the world of playing, there is a growing interest in spectating digital games \cite{Egliston_2020b}. Researchers have studied the reasons behind its growth \cite{Taylor_2018,Woodcock_Johnson_2019}, the types of game spectators \cite{Golob_Krasevec_Crnic_2021}, and the roles they take on while spectating \cite{Lin_Sun_2011}. Spectatorship allows people to form social connections and communities by sharing opinions and spaces and learning new strategies to improve their skills \cite{Egliston_2020b}. However, the traditional forms of spectating are ``vicarious'', meaning that spectators enjoy playing ``through'' others rather than themselves \cite{Golob_Krasevec_Crnic_2021}. These vicarious experiences do not engage the spectator's body in the gameplay in any way, limiting their engagement to only observing the person playing \cite{Orme_2021}.

\begin{figure}[h]
  \centering
  \includegraphics[width=\textwidth]{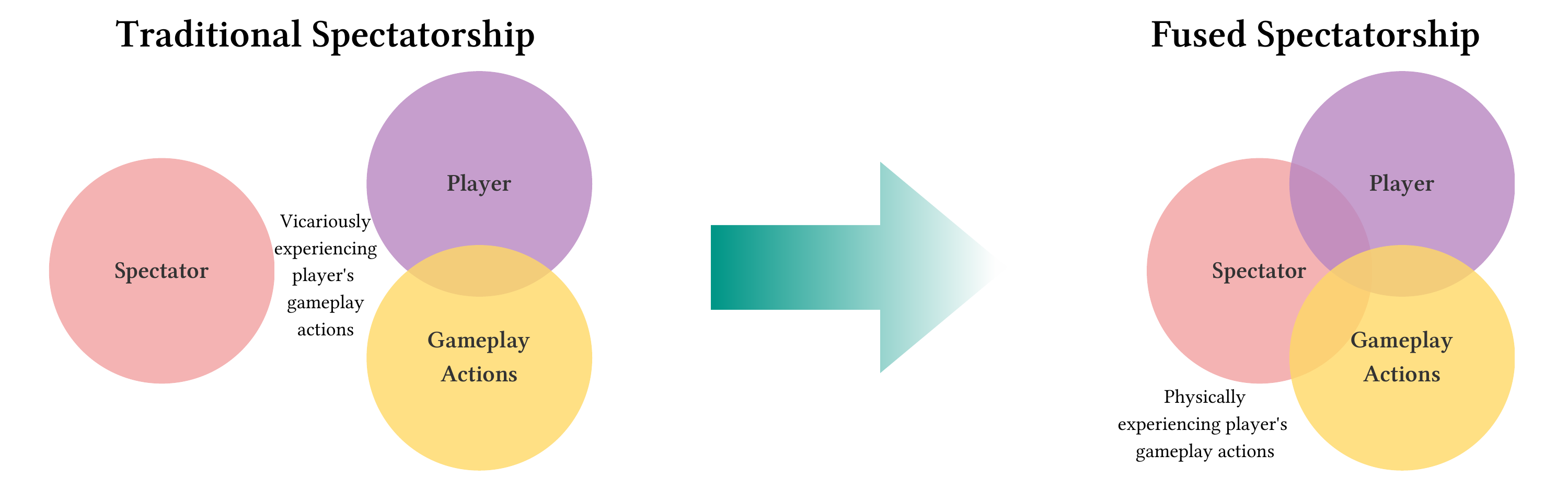}
  \caption{A Venn diagram showing the relationship between the spectator, player and gameplay actions in traditional game spectatorship \cite{Golob_Krasevec_Crnic_2021} and Fused Spectatorship.}
  \Description{A Venn diagram showing the relationship between the spectator, player and gameplay actions in traditional game spectatorship and Fused Spectatorship.}
  \label{fig:teaser}
\end{figure}

For enhancing affective engagement with digital games \cite{Bianchi-Berthouze_Kim_Patel_2007}, the Human-Computer Interaction (HCI) design community has been using advancements in body-centric technologies \cite{Mueller_Andres_Marshall_Svanas_schraefel_Gerling_Tholander_2018} to explore ways to make spectating a more bodily experience \cite{Ludvigsen_Veerasawmy_2010,Tekin_Reeves_2017}. For example, researchers designed spectatorship systems to support interaction through emotional signals \cite{Khot_Andres_Lai_von_Kaenel_Mueller_2016} or claps \cite{Tomitsch_Aigner_Grechenig_2007}, to make them feel like they are more actively involved in the gameplay. However, current attempts at bodily spectatorship still rely heavily on screens. This might limit the spectators' urge to be more physically involved to have a meaningful engagement with the gameplay \cite{Seering_Savage_Eagle_Churchin_Moeller_Bigham_Hammer_2017}. We believe that more physical engagement is now possible with actuation technologies such as Electrical Muscle Stimulation (EMS) that can use the human body as an output medium \cite{Patibanda_Li_Chen_Saini_Hill_2021,Patibanda_Van_Mueller_2022,Patibanda_Semertzidis_Scary_La_2020,Nisal_Patibanda_2022}.

\begin{figure}[h]
    \centering
    \includegraphics[width=1\textwidth]{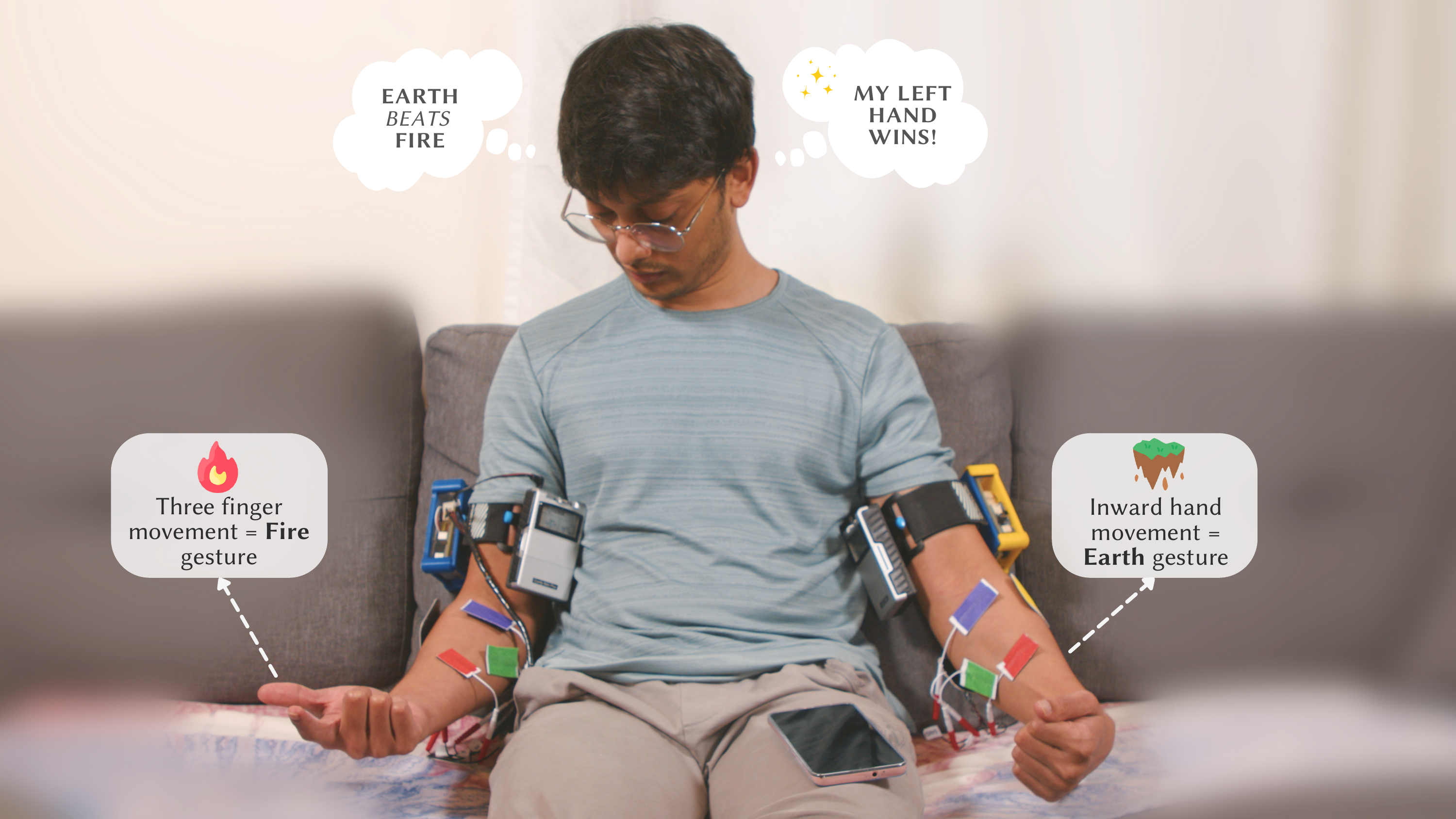}
    \caption{A participant spectates their EMS-controlled hands playing the Godai game (similar to rock-paper-scissors).}
    \Description{A participant spectates their EMS-controlled hands playing the Godai game (similar to rock-paper-scissors).}
    \label{fig:introgodaiplaying}
\end{figure}

We propose a new approach - \textit{Fused Spectatorship} - where a spectator's body is more actively engaged in spectating gameplay by watching their own body play games. The term ``Fused'' is apt for our work as it connotes the integration of the spectator's body in the physical world (from which they are spectating the gameplay) with the player's digital world (Fig. \ref{fig:teaser}). To showcase this approach, we designed three games - similar to traditional hand games like rock-paper-scissors - where a person loans control over both their hands to a computational EMS system \cite{Knibbe_Alsmith_Hornbak_2018,Lopes_Baudisch_2017,mueller2023towards,Floyd_Mueller_Patibanda_Byrne_Li_Wang_Andres_2021} to spectate their (EMS-controlled) hands play these games (Fig. \ref{fig:introgodaiplaying}).

We obtained institutional ethics approval to conduct an in-the-wild study \cite{Rogers_Marshall_2017} with 12 participants and interviewed them using a semi-structured approach \cite{Longhurst_2003}. We analysed the interview data using an inductive thematic analysis approach \cite{Braun_Clarke_2006} to articulate four user experience themes. By analysing the themes, we discovered four fused spectator types. We discuss the generalised behaviour the four spectator types elicited and offer one design consideration each to support this behaviour. Our work helps reflect on the coming together of spectatorship \cite{Taylor_2018,Woodcock_Johnson_2019}, bodily games \cite{Salen_Zimmerman_2004,Li_Captcha_Chen_Patibanda_Mueller_2021} and bodily control \cite{Mueller_Petersen_Li_2023,Mueller_Lopes_Strohmeier_Ju_Seim_Weigel_Nanayakkara_2020} to understand: 1) the experience of using the physical body for spectating in digital games; 2) how playing and spectating can come to be intertwined; and, 3) the potential and limitations of our approach to move beyond the traditional 2D screen player-spectator interactions. Our work makes the following contributions:

\begin{itemize}
    \item A system contribution by presenting three games, adding to HCI's collection of novel systems \cite{Wobbrock_Kientz_2016}. This contribution can inspire developers in industry to utilise EMS through the lens of entertainment to create engaging experiences benefiting, for example, motor-rehabilitation practices.
    \item An empirical contribution by presenting results as four user experience themes. This contribution can be useful for user experience researchers who aim to understand the potential and limitations of using \textit{our} approach.
    \item A theoretical contribution by presenting four fused spectator types to extend existing spectator types when watching digital games over a screen \cite{Golob_Krasevec_Crnic_2021}. We also articulate the generalised behaviour of each spectator type and offer one design consideration for supporting it, which can be useful for practitioners interested in creating fused spectatorship experiences.
\end{itemize}

\section{RELATED WORK}\label{sec:RW}
Our work is inspired and informed by the change in spectatorship along mass media consumption, the vicarious nature of digital game spectatorship, developments towards engaging the body in spectatorship and using interactive technology turn the body into a display, especially by using Electrical Muscle Stimulation (EMS).

\subsection{Change in Spectatorship Along Mass Media Consumption}\label{sec2.1:specagency}
As we trace the journey of mass media, we can see a shift in how spectators engage with content. The invention of the radio in the early 20th century marked a turning point in mass communication, enabling people to listen to news, music, and stories from the comfort of their homes \cite{Douglas_2004}. Here, we note that radios provided a passive shared auditory spectating experience. This means that people listening to radios cannot actually see anything; however, the audio provided them a means to passively engage with the content through their imagination by creating mental images based on the auditory cues \cite{Radio_Voices}. The cinema brought visual storytelling to the masses, allowing audiences to immerse themselves in a narrative through moving images \cite{Gunning_2004}. Although still largely passive, this expanded the possibilities for emotional and sensory engagement, as visual elements played a crucial role in the spectator's experience \cite{Bordwell_Thompson_Smith_2008}. With advancements in digital technology and taking inspiration from game genres such as role playing games~\cite{Li_VR_Tang_Tong_Patibanda_Mueller_Liang_2021}, interactive movies emerged, allowing audiences to influence the narrative by making choices for the characters \cite{Vosmeer_Schouten_2014}. For example, Netflix offers interactive movies such as ``Black Mirror: Bandersnatch'' \cite{Elnahla_2020_blackmirror}. These interactive movies bridged the gap between passive watching and active participation, hoping to foster a more engaging and immersive experience for the audience. This change in spectatorship through mass media developments inspired us to think ahead of where bodily digital games could go, possibly allowing for more engaging forms of spectating. For this, we also learned from the concept of spectatorship and its vicarious nature in the context of digital games, which we discuss next.

\subsection{The Vicarious Nature of Digital Game Spectatorship}\label{sec2.2:vicariousnatureofspec}
Spectatorship refers to ``the experience of being a spectator'', while spectating refers to ``the act of observing or watching something by a spectator'' \cite{Mayne_1993}. Digital game spectatorship has been gaining popularity since the early arcade days, where spectators gathered around to watch live gameplay \cite{McClelland_2006,Taylor_2016}, and researchers are beginning to understand the increasing trend around spectating digital games \cite{Egliston_2020a,Gandolfi_2016,Scully-Blaker_Begy_Consalvo_Ganzon_2017,Woodcock_Johnson_2019}. For example, researchers proposed various roles spectators take on when spectating digital games and found that some like to be more involved than others \cite{Lin_Sun_2011}. They have also identified five types of spectators: spectator, performer, selector, viewer, and substitutor \cite{Golob_Krasevec_Crnic_2021}, suggesting that people enjoy spectating digital games ``vicariously'' as their bodies are passively engaged and not necessarily performing the act of playing \cite{Cheung_Huang_2011}. Burwell \cite{Burwell_2017}, on the other hand, suggests that spectators actively participate in the meaning-making of the game \cite{Burwell_Miller_2016} and enjoy the vicarious play that unfolds in their minds parallel to the active play of the player \cite{Cheung_Huang_2011}. These works suggest that spectators can actively engage cognitively and affectively, but these works also highlight that the physical body is still not involved in spectatorship. \textit{What, then, are the spectator types that emerge when spectating digital games by actively engaging the physical body in spectatorship?} We answer this question through our work in section \ref{sec6:fusedspectatortypes}.

Prior work has also tried to understand the bodily aspects involved in spectatorship. For example, when spectators of esports attempt to emulate the pro's play using strategies and techniques derived from spectatorship, they realise quickly that they usually cannot fully perform all the bodily actions they saw, highlighting that ``seeing isn't doing'' \cite{Egliston_2020a}. Scully-Blaker et al. \cite{Scully-Blaker_Begy_Consalvo_Ganzon_2017} also described this tension between the spectator's desire to watch and learn and the esports player's duty to play while entertaining the crowd. This work highlights the bodily labour that is often invisible until they reach a certain level of popularity \cite{Woodcock_Johnson_2019}. These works suggest an opportunity to tighten the relationship between the players and spectators. Recognising this opportunity, streaming platforms increasingly offer features such as a live chat \cite{Striner_Webb_Hammer_Cook_2021}. However, these features are still, so far, mostly confined to the screen and therefore miss an opportunity to make use of the affordances of more embodied approaches to interaction design \cite{Dourish_2001}. The next section presents what we learned from prior work around how emerging HCI design engaged the body in spectatorship.

\subsection{Towards Engaging the Body in Spectatorship}\label{sec:engagingthebodyspec}
The game design community in HCI has begun to recognise spectators' importance to the experience of play \cite{Cheung_Huang_2011}. In response, researchers have designed systems that use the spectator's body as input to influence their experience. For instance, Tomitsch et al. \cite{Tomitsch_Aigner_Grechenig_2007} designed motion-controlled wearables for sports that allows spectators to take up the role of judging performances. From this work, we learned that wearables could support embodied forms of spectating, and thus, we focused on developing a wearable in our design work. However, it is unclear how to actively engage the spectator's body to spectate digital games since previous research has focused on spectating sports.

Other researchers have explored the spectrum of spectatorship experiences through a broader framework \cite{Reeves_Benford_OMalley_Fraser_2005}. The framework identifies various spectating experiences, from simple photo booth experiences to complex performances involving exoskeletons and proposes design strategies for these experiences. We find the ``magical'' strategy intriguing, where the performer's actions are ``magical'' as technology reveals their effects to the spectators but hides the actions that caused them. EMS technology seems to lend itself nicely to such ``magical'' spectator experiences, as research shows that users can find the technology ``weirdly'' intriguing \cite{Knibbe_Alsmith_Hornbak_2018,Lopes_Ion_Mueller_Hoffmann_Jonell_Baudisch_2015}. Hence, we started our investigation with EMS technology, although we acknowledge that other technologies, such as exoskeletons, are possible.

\subsection{Using Interactive Technology to Turn the Body Into a Display}\label{sec2.3:bodyasdisplay}
Prior work has highlighted that engaging the physical body is important for engagement and amplifying the affective experiences of players \cite{Bianchi-Berthouze_Kim_Patel_2007,Li_Huang_Patibanda_Mueller_2023}. Particularly, Ludvigsen and Veerasawmy \cite{Ludvigsen_Veerasawmy_2010} suggest that, when it comes to game spectatorship, designers should strive to intimately connect the spectators and players \cite{Sigman_others_2007}. Hence, we turned to prior work that aimed to engage the body, in particular, turned the body into a display.

\subsubsection{Using the Body as a Passive Visual Display}\label{sec2.3.1:passivedisplay}
HCI researchers have utilised interactive technology to actively engage the body by transforming its surface into an interactive visual display. For instance, the ``body landmarks'' system turns specific body parts, such as knuckles, into an interactive display, prompting people to shift their focus to those areas \cite{Steimle_Bergstrom_Weigel_Nittala_Boring_Olwal_Hornbak_2017}. Moreover, augmented reality systems allow the body to be turned into a visual display, such as ``DigiGlo'' \cite{Chatain_2020}, localising the visuals in the player's palms and ``mirracle'' \cite{Blum_Kleeberger_Bichlmeier_Navab_2012} throughout the entire body. These works illustrate that the body can be used either as a localised display or as a whole-body display. However, while necessitating the involvement of the user's body, these technologies still allow them to retain complete control over their bodily movements. This implies that the body is still used as a passive display. Therefore, we turn to EMS technology, as it could convert the body into an active visual display by utilising it as an output medium.

\subsubsection{Using EMS to Turn the Body Into an Active Visual Display}\label{sec2.3.2:EMSdisplay}
Electrical Muscle Stimulation (EMS) can use the body as an output medium by passing a small amount of electric current via electrodes attached to the body to create involuntary bodily movements \cite{Knibbe_Alsmith_Hornbak_2018}. Previous research has used EMS to create experiences that use the body as an active visual display. For example, Muscle Plotter \cite{Lopes_Yuksel_Guimbretiere_Baudisch_2016} is a system that actuates subtle hand movements that are invisible to onlookers, allowing users to sketch industrial designs such as car aerodynamics. Another system created by Pfeiffer et al. \cite{Pfeiffer_Dunte_Schneegass_Alt_Rohs_2015} moves the user's foot by actuating involuntary gross motor movements that are visible to observers. These works taught us that EMS could actuate fine- and gross-motor movements. Therefore, we incorporated both types of movements to engage the spectator's body as an active display in the fused spectatorship experience. However, precise hand and finger movement control with EMS alone can be difficult to achieve \cite{Nith_Teng_Li_Tao_Lopes_2021}. Therefore, we focused on designing games such as rock-paper-scissors that do not rely on the execution of specific hand and finger movements. Moreover, this work does not provide insight into the user's experience of spectating the body performing EMS-controlled gestures, which our study addresses.

In summary, we hypothesise that technologies like EMS allow for ``magical'' spectator experiences where people loan bodily control to the computer \cite{Reeves_Benford_OMalley_Fraser_2005}. This suggests a potential to engage the human body more actively by using it as an output medium. Nonetheless, there is still little comprehension of how actuation technology's capability could intertwine the spectator-player relationship in digital games spectatorship. Therefore, this article attempts to answer the research question: \textit{How do we design spectating experiences where the spectator's body is more actively engaged in the gameplay?}

\section{DESIGNING THE GAMES FOR FUSED SPECTATORSHIP}\label{sec3:gamesdesigns}
To answer our research question, we propose \textit{Fused Spectatorship} – where the spectator's physical body is more actively engaged in spectating digital games. In this section, we articulate the design of our EMS system (Fig. \ref{fig:participantgamekit}) and the three games. The games are called, ``Godai'' and ``Ept\'{a}'' - where the hands compete, and ``\'{I}dio'' – where the hands collaborate towards a common goal (explained in section (Fig. \ref{sec3.5:threegames}).

\subsection{The Computational EMS System}\label{sec3.1:EMSsystemdesign}
EMS interactions may feel uncomfortable at first \cite{Knibbe_Alsmith_Hornbak_2018}; therefore, loaning bodily control to the EMS system requires the person to trust its capabilities and know the limitations \cite{Wagner_Robinette_Howard_2018}. To build trust, we designed hardware and software features that aim to inform participants about what the system can do, provide feedback on its behaviour, and ensure participants can take back control if necessary.

\subsubsection{Hardware}\label{sec3.1.1:hardware}
\begin{figure}[hbt!]
    \centering
    \includegraphics[width=1\textwidth]{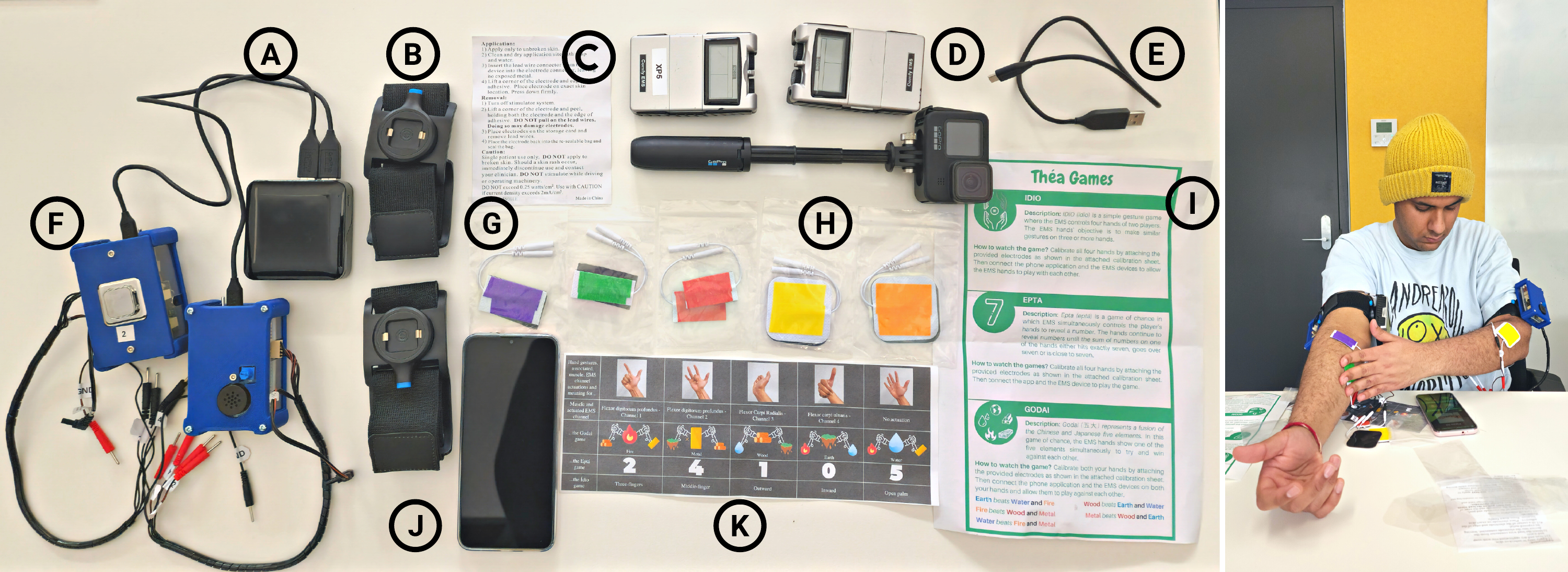}
    \caption{(Left) shows the computational EMS system provided to the participant for the study. (A) cables and power bank, (B) two Velcro arm bands, (C) EMS safety instructions, (D) two EMS devices, (E) additional charging cable, (F) microcontroller devices, (G) colour-coded electrodes, (H) GoPro camera, (I) games description leaflet, (J) phone with a software application installed, and (K) instruction sheet showing hand-gestures and associated meanings for the games. (Right) shows a participant calibrating using instructions on (J) during the pre-study session.}
    \Description{(Left) shows the computational EMS system provided to the participant for the study. (A) cables and power bank, (B) two Velcro arm bands, (C) EMS safety instructions, (D) two EMS devices, (E) additional charging cable, (F) microcontroller devices, (G) colour-coded electrodes, (H) GoPro camera, (I) games description leaflet, (J) phone with a software application installed, and (K) instruction sheet showing hand-gestures and associated meanings for the games. (Right) shows a participant calibrating using instructions on (J) during the pre-study session.}
    \label{fig:participantgamekit}
\end{figure}

The EMS system's hardware (Fig. \ref{fig:participantgamekit}) includes a mobile phone with pre-installed software (same for all games), two EMS devices, Bluetooth-enabled microcontroller devices (placed in a 3D printed case with a speaker), Velcro bands, and electrodes. The EMS devices and microcontrollers are attached to the Velcro bands and then to the spectator's arms (approximately 350 grams). The microcontroller device splits one channel of the EMS device into four controllable EMS channels. A physical copy of instructions to calibrate the electrodes was provided to complement the software's instructions. To ensure safety, we used a commercial EMS device whose parameters like intensity, pulse rate, and width could not be programmatically controlled, as prior work suggested \cite{Knibbe_Alsmith_Hornbak_2018,Kono_Takahashi_Nakamura_Miyaki_Rekimoto_2018,Lopes_Ion_Mueller_Hoffmann_Jonell_Baudisch_2015}. Moreover, the microcontroller has a physical switch to override the software and turn off the stimulation if needed. The electrodes were colour-coded to indicate the gestures they could actuate, making the calibration process quicker. We also provided participants with a GoPro action camera to record themselves at home (section \ref{sec4:studydesign}).

\subsubsection{Software}\label{sec3.1.2:software}
We developed an Android application that uses Bluetooth to communicate with the microcontrollers and trigger the correct electrode combinations for each game. The app also tracks the participants' nicknames and logs the details of their engagement, such as how many times and for how long they spectated. Participants can choose from different game modes, including best of 3, 5, or free play. While Godai has all three modes, Ídio and Épta only offer free play (game details are discussed in section \ref{sec3.4:preparing}). The software provides clear instructions for calibrating the electrodes and visually guides users on electrode placement, which is the same for all games. We reused the electrode positioning in all three games to eliminate the need for recalibration, which can be a tedious process \cite{Gange_Knibbe_2021}.

\begin{figure}[hbt!]
    \centering
    \includegraphics[width=1\textwidth]{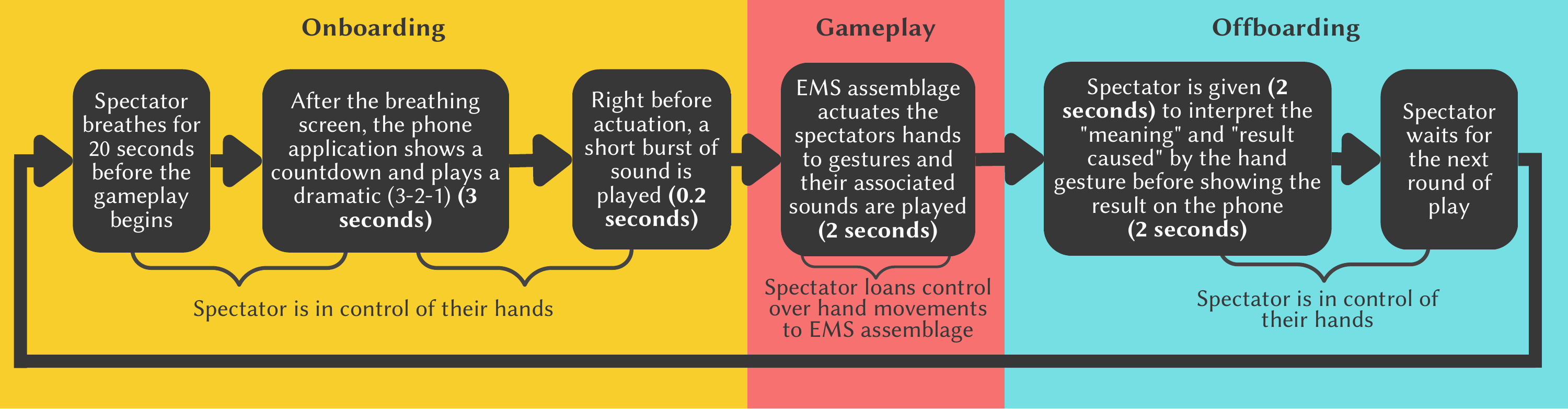}
    \caption{Control loop.}
    \Description{Control loop.}
    \label{fig:controlloop}
\end{figure}

\subsection{Software Architecture: The control loop}\label{sec3.2:controlloop}
We created a software architecture to govern how the EMS system interacts with the body during each round of play - we call it the ``control loop'' (Fig. \ref{fig:controlloop}). The control loop is divided into three phases: onboarding, gameplay, and offboarding. These phases have overlapping features that inform the participant of the EMS system's capabilities, reveal its behaviour, and allow to take back control immediately.

\subsubsection{Onboarding Phase}\label{sec3.2.1:onboarding}
This phase has two software features.

\textbf{Breathing screen:} The breathing screen appears after clicking the ``Play'' button on the calibration screen, with a skip button at the bottom. In the centre, we designed a circular animation of a flower that slowly expands and contracts, with a text message asking the spectator to breathe in sync. The design feature aims to help participants relax their body and muscles, aiding in reducing the tingling and uncomfortable sensations caused by EMS. We believe that relaxing the body also allows the EMS to take better control over their body, creating more accurate muscle actuations.

\textbf{EMS countdown:} After the breathing screen, an on-screen 3-2-1 countdown is visually displayed, accompanied by a dramatic sound played in sync on the in-game screen. This safety feature was designed to prepare the spectator's body for the EMS actuation. By playing this sound, our aim was to reduce any potential discomfort and enhance the overall experience of the spectator, making it more enjoyable and engaging.

\subsubsection{Gameplay Phase}\label{sec3.2.2:gamplayphase}
This phase has three software features.

\textbf{Sound feedback:} A sound feedback feature was provided after the countdown. The participant can choose from three options: 1) Turn the two-pitched sound on, 2) turn the second pitch off, and 3) turn the sound off completely. With option 1, the first pitch (not available for option 3) is played immediately before the EMS takes control of the hands to help participants know when to loan bodily control to the system. The second pitch (not available for options 2 and 3) differs for each of the five gestures.

\textbf{Actuation of the gesture:} The software installed on the individual microcontroller devices triggers an EMS channel, which then actuates a hand gesture for that round of spectatorship~\cite{Li2022gesplayer}. To ensure the safety of the participants, we actuated their hands only for a short time, i.e., two seconds in our case.

\textbf{Voice control:} We included a voice control feature to give participants control over the EMS actuations. A participant can pause and resume the game by saying ``stop'', ``pause'', or ``resume''.

\subsubsection{Offboarding Phase}\label{sec3.2.3:offboarding}
This phase has only one software feature called the \textbf{Reveal button}. Our goal was to make participants focus on their body movements. Therefore, the design default was for results to be hidden behind a reveal button on the in-game screen. If participants chose to see the screen while spectating their body, they could click the reveal button, which shows the results for two seconds. This window of opportunity to look at the screen was designed to help participants interpret the meaning of the actuated hand gesture if the EMS could not fully actuate their hands, possibly due to the change in their body orientation while playing.

\begin{table}[]
\caption{Key characteristics of the three games.}
\label{tab:keygamechars}
\centering
\resizebox{\textwidth}{!}{%
\begin{tabular}{lllll}
\toprule
\textbf{Théa game} & \textbf{Physical engagement} & \textbf{Type of motor-movement} & \textbf{Game outcome} & \textbf{Play dynamics} \\
\toprule
Godai & Two hands (max) & Fine- and gross motor-movement & Symbolic  & Competitive \\
\midrule
Eptá  & Two hands (max) & Fine- and gross motor-movement & Numerical & Competitive \\
\midrule
Ídio  & Two hands (max) & Fine- and gross motor-movement & Symbolic  & Collaborative \\
\bottomrule
\end{tabular}%
}
\end{table}

\subsection{The Three Games}\label{sec3.3:threegames}
We describe the design of our games using four characteristics: 1) physical engagement \cite{Mueller_Gibbs_Vetere_2008}, 2) motor-movement \cite{Mueller_Agamanolis_Picard_2003}, 3) game outcome \cite{Isbister_Mueller_2015} and 4) play dynamics \cite{Skultety_2011}, previously identified as key to designing digital bodily games \cite{Jensen_Rasmussen_Gronbak_2013} (Table \ref{tab:keygamechars}).

\begin{itemize}
    \item \textit{Physical engagement} - what body parts does the participant loan bodily control over?
    \item \textit{Motor-movement} – what type of motor movement is controlled by the computer?
    \item \textit{Game outcome} – what type of game outcome results from loaning control?
    \item \textit{Play dynamics} – what are the dynamics emerging from loaning control?
\end{itemize}

\subsection{Preparing to Spectate the Games}\label{sec3.4:preparing}
\begin{figure}[hbt!]
    \centering
    \includegraphics[width=1\textwidth]{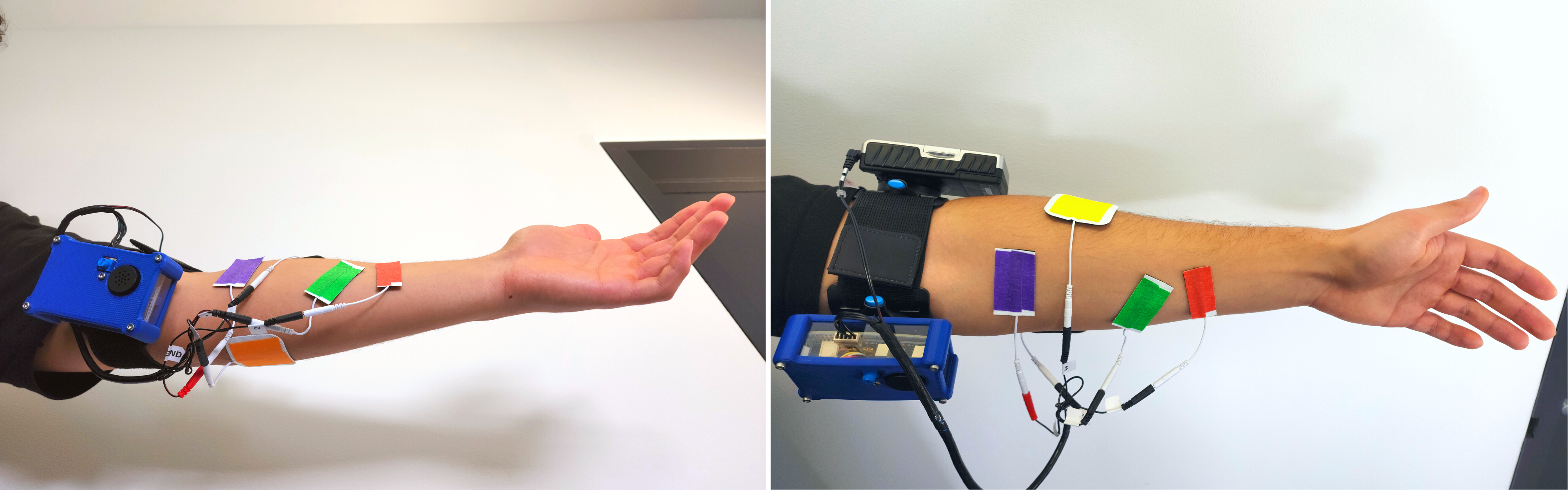}
    \caption{(Left) A bottom and top (Right) view of a participant's forearm showing the electrode placement.}
    \Description{(Left) A bottom and top (Right) view of a participant's forearm showing the electrode placement.}
    \label{fig:electrodeplacement}
\end{figure}

\begin{figure}[hbt!]
    \centering
    \includegraphics[width=1\textwidth]{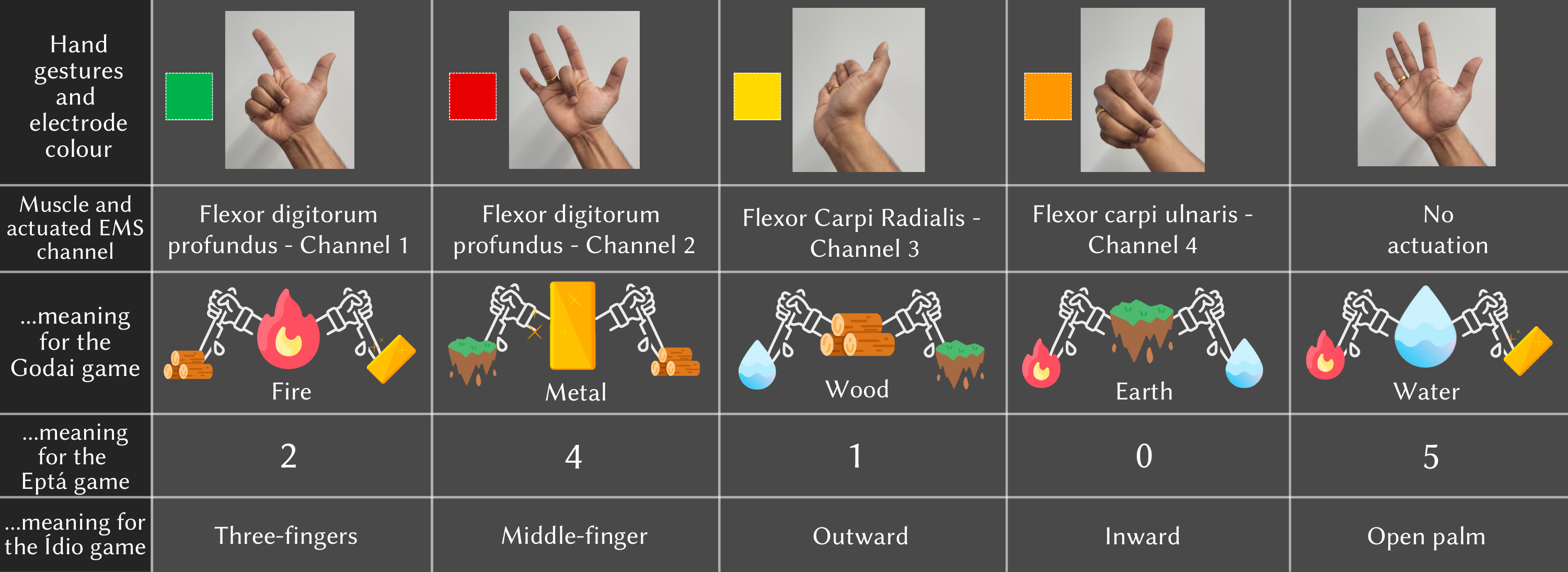}
    \caption{Each hand gestures, EMS channel numbers that actuate them and their meanings in each game.}
    \Description{Each hand gestures, EMS channel numbers that actuate them and their meanings in each game.}
    \label{fig:gamehandgestures}
\end{figure}

The participant wears the hardware on their arms (Fig. \ref{fig:electrodeplacement}). They connect the EMS devices using the software application, choose the game and mode of play, calibrate the electrodes (attached to four EMS channels) to achieve the required game movements, and click on ``play''. This action triggers the control loop described in section \ref{sec3.2:controlloop}. The EMS-controlled gestures are the same in all games. However, their associated meaning differs for each game (shown in Fig. \ref{fig:gamehandgestures}). We now individually describe each game.

\subsection{Design of the Three Games}\label{sec3.5:threegames}
We took inspiration from traditional hand games like rock-paper-scissors and designed three games. We call them ``The\'{a}'' games, Greek for ``watching''. These games can be spectated using the body as an output medium by wearing our EMS system on both hands as a single spectator.

\subsubsection{Game 1: Godai}\label{sec3.5.1:godai}
\begin{figure}[hbt!]
    \centering
    \includegraphics[width=1\textwidth]{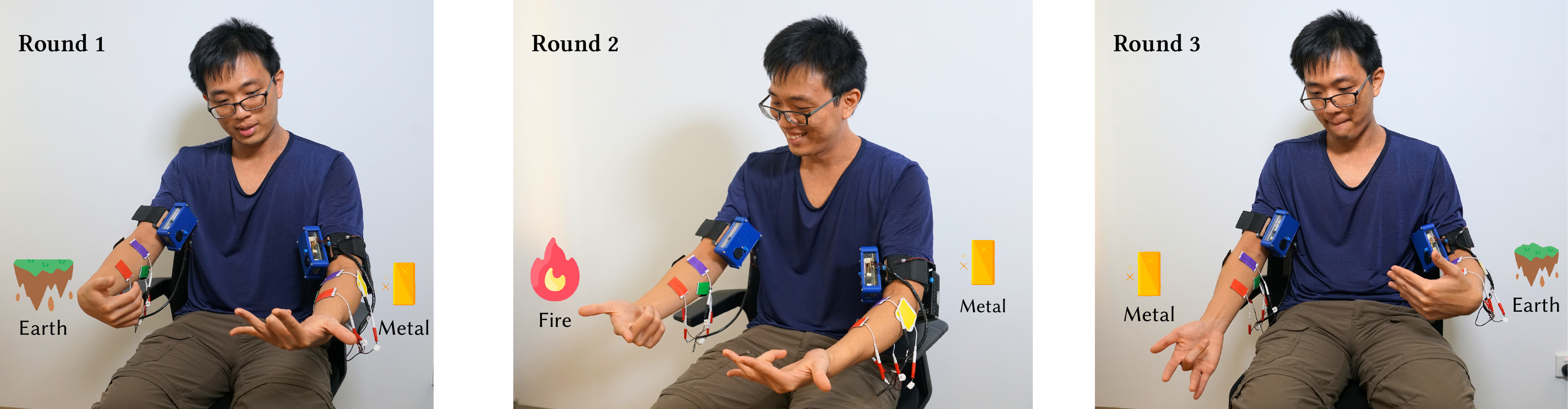}
    \caption{Three rounds of the Godai game.}
    \Description{Three rounds of the Godai game.}
    \label{fig:godai}
\end{figure}

Godai \begin{CJK*}{UTF8}{min}(五行)\end{CJK*} represents a fusion of the Chinese \cite{Wuxing_Chinese_philosophy_2022} and Japanese \cite{Godai_Japanese_philosophy_2022} five elements. In this game, the EMS-controlled hands simultaneously display one of the five elements shown in Fig. \ref{fig:gamehandgestures}. Fig. \ref{fig:godai} illustrates a spectator experiencing the best of three rounds of this game. In the left image, Round 1, the right hand displays "Earth", while the left hand displays "Metal". Since "Metal" beats "Earth", the left hand wins this round and earns the point. Similarly, in the centre image, Round 2, the right hand displays "Fire", and the left-hand displays "Metal". Since "Fire" beats "Metal", the right hand wins this round and earns the point. Finally, in the right image, Round 3, the right-hand displays "Metal", and the left-hand shows "Earth". Since "Metal" beats "Earth", the right hand wins this round and earns the point, ultimately defeating the left hand with a score of 2 to 1. The dynamic of this game probably results in the person feeling competitive as the EMS-controlled hands display gestures to outdo each other.

\subsubsection{Game 2: Ept\'{a}}\label{sec3.5.1:epta}
Ept\'{a} ("seven" in Greek) is a game of chance inspired by Blackjack \cite{Blackjack_2022}. Both hands take turns showing a number (Fig. \ref{fig:gamehandgestures}). This turn-by-turn mechanism mimics the card dealer dealing cards in traditional Blackjack. The hands continue to reveal numbers until the sum of numbers shown by one hand is exactly seven, which makes that hand win the round. Otherwise, if this hand goes over seven, it loses the game. The dynamic of this game probably results in the participant feeling a sense of competition.

\begin{figure}[hbt!]
    \centering
    \includegraphics[width=1\textwidth]{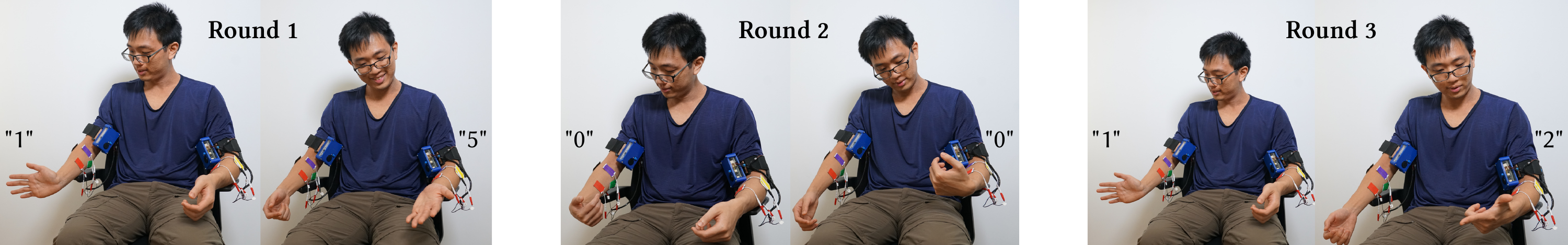}
    \caption{Three rounds of the Ept\'{a} game.}
    \Description{Three rounds of the Ept\'{a} game.}
    \label{fig:epta}
\end{figure}

For example, Fig. \ref{fig:epta} shows a spectator experiencing three rounds of a game of Ept\'{a}. In the leftmost image - Round 1, the right hand displays a gesture representing "1", and the left hand represents "5". In the centre image - Round 2, the right hand reveals a gesture representing "0", and the left hand represents "0". In the rightmost image - Round 3, the right hand shows a gesture representing "1", and the left hand represents "2". Since the sum of numbers on the right hand is (1 + 0 + 1 = 2), and the left hand is (5 + 0 + 2 = 7), the left hand wins this game as it shows numbers whose sum is exactly "7".

\subsubsection{Game 3: \'{I}dio}\label{sec3.5.1:idio}
\'{I}dio (Greek for `same') is a simple gesture game inspired by Mahjong \cite{Mahjong_2022}, a tile-based game. In Mahjong, players must form four sets and one pair of the same tiles, and the tiles are visible. In \'{I}dio, both the EMS-controlled hands randomly show one of the five gestures shown in Fig. \ref{fig:gamehandgestures}. Here, the "hand gestures" are the "tiles", and they are invisible to the EMS hands, i.e., the EMS devices are incapable of predicting the next or learning from the previous hand gesture. When three or more hands show the same gesture, it is struck off on the phone application and is not shown by the EMS hands again. These hands continue to show the gestures until they show all five gestures, marking the end of the game. The dynamic of this game results in the spectator feeling collaborative as both their hands are trying to show the same gesture to complete the game.

\begin{figure}[hbt!]
    \centering
    \includegraphics[width=1\textwidth]{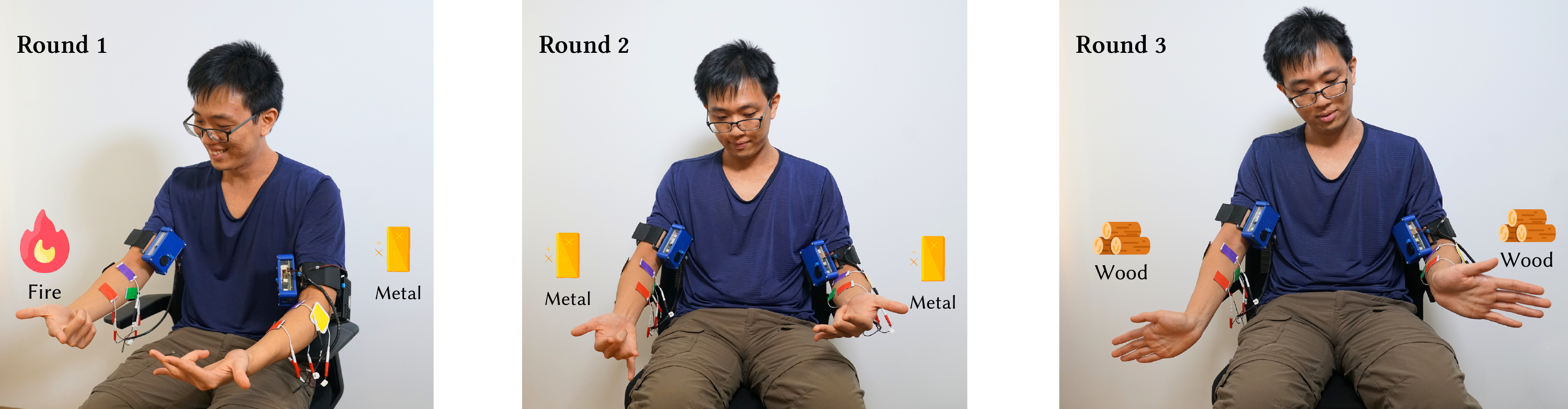}
    \caption{The first three rounds of a game of \'{I}dio.}
    \Description{The first three rounds of a game of \'{I}dio.}
    \label{fig:idio}
\end{figure}

For example, Fig. \ref{fig:idio} shows a spectator experiencing the first three rounds of the \'{I}dio game. In Round 1, the right hand shows a gesture representing "Fire," and the left hand shows "Metal". Since the gestures are different, the game proceeds to the next round. In Round 2, the right hand shows a gesture representing "Metal", and the left hand shows "Metal". Similarly, in Round 3, the right hand shows a gesture representing "Wood", and the left hand shows "Wood". As both hands show the same gesture in Rounds 2 and 3, these gestures are marked off on the phone application and will not be shown by the EMS-controlled hands again. The game continues in this manner until all five gestures have been shown by both hands, marking the end of the game.

\section{STUDY DESIGN AND DATA ANALYSIS}\label{sec4:studydesign}
\begin{figure}[hbt!]
    \centering
    \includegraphics[width=1\textwidth]{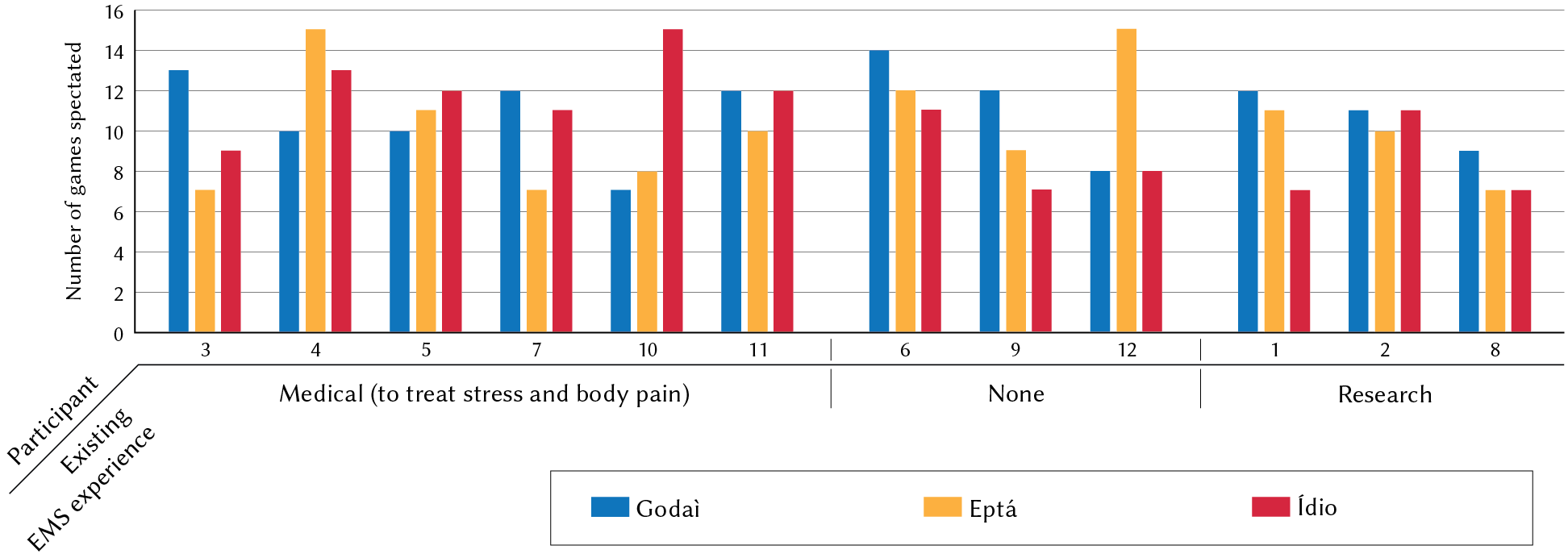}
    \caption{Grouping of participants based on their experience with EMS.}
    \Description{Grouping of participants based on their experience with EMS.}
    \label{fig:demographics}
\end{figure}

To gain insight into the user experience of engaging with our games, we obtained institutional ethics approval to conduct an in-the-wild study \cite{Chamberlain_Crabtree_Rodden_Jones_Rogers_2012,Rogers_Marshall_2017}. This study method, as demonstrated by previous work on novel game experiences \cite{Semertzidis_Scary_Andres_Dwivedi_Kulwe_Zambetta_Mueller_2020}, allows researchers to collect rich data from participants and reduces the potential impact of any personal biases of the researchers. In our case, this method was beneficial as it allowed participants to engage with our system without researchers present and in a comfortable environment of their choice, such as their homes, and during their preferred times. For example, we designed the system to understand a single spectator's experiences. However, through our in-the-wild study, we learnt that two individual spectators could use our system's two sets of EMS devices for a shared fused spectatorship experience. Here, a spectator can choose to wear only one device on one arm and give another device to another person (Fig. \ref{fig:participantlookingatphone}). This shows the value of in-the-wild research where participants are free to use the designed system to suit their needs rather than being constrained by pre-established notions that researchers might have imposed \cite{Lopes_Ion_Mueller_Hoffmann_Jonell_Baudisch_2015,Reeves_Benford_OMalley_Fraser_2005}.

We recruited 12 participants, of which six had prior experience using an EMS device to treat body stress or pain, three used EMS for research, and three had no experience with EMS. The participants had a mean age of 30.9 years, with a standard deviation (SD) of 6.52. We describe the experiences of these participants in section \ref{sec5:UXthemes}, and Fig. \ref{fig:demographics} shows the number of participants and the number of times they played each game.

As described in section \ref{sec3:gamesdesigns}, we used the principles of capability, feedback and safety principles \cite{Kono_Takahashi_Nakamura_Miyaki_Rekimoto_2018,Wagner_Robinette_Howard_2018} for guiding our study design. We believe these principles helped us build trust in our participants in relation to engaging with a system that takes control over their body. We engaged with the principles across three study phases: a pre-study, an in-the-wild study, and a post-study interview, which we describe next.

\subsection{Pre-study Phase: Safe onboarding and informing system's capability} \label{sec4.1:prestudyphase}
The pre-study phase's goal was to introduce the research topic to participants, demonstrate the capabilities of the EMS system, and allow them to safely try it out with the research team present. We used this phase to show participants how to use all components of the EMS system, calibrate the electrodes following the instructions on the smartphone app, and experience the games. Participants were also shown how to use the voice control. Then, participants followed the procedure independently, with the researchers available to answer any questions. After calibration, participants spectated the games at least twice under the research team's supervision before taking the EMS system home. We also aimed to understand the order of game difficulty. We explored this by asking six participants to watch the games in the order of \'{I}dio, Ept\'{a}, and Godai and reversed the order for the other six participants. The average time for the pre-study phase was 51 minutes, with SD = 13 minutes.

\subsection{In-the-wild Phase: Feedback, study and data collection}\label{sec4.2:wildphase}
During the in-the-wild phase, participants were given access to the researchers via phone and email during work hours and were encouraged to contact them with any questions or feedback. Participants had the EMS system for one day and were asked to spectate each game at least seven times. Participants were also asked to record themselves using the GoPro camera we provided along with the system while spectating their EMS-controlled hands play the games (as described in section \ref{sec3.1:EMSsystemdesign}). While participants were encouraged to engage with the EMS system alone, eight participants experienced the system by also sharing the system with a friend, partner, or family member. In these cases, two participants calibrated one hand each to spectate the games with each other. While the participants experienced the games socially, we conducted individual interviews to analyse their experiences. We collected the EMS system from the participants the next day, concluding this study phase.

\subsection{Post-study Phase: Further data collection and data analysis}\label{sec4.3:poststudyphase}
During the post-study phase, we interviewed participants for approximately 90 minutes, recording audio and video to understand their experiences (mean duration = 74.38 minutes, SD = 14.35). The interviews were semi-structured \cite{Dearnley_2005}, and we used the laddering technique \cite{Trocchia_Swanson_Orlitzky_2007}. After transcription, two coders independently conducted an inductive thematic analysis \cite{Braun_Clarke_2006} of the interviews using NVivo software \cite{Wiltshier_2011}. Each answer was considered as one data unit, with an average word count of 64. The coders assigned each data unit a category code, creating 236 and 197 initial codes. After refinement, we arrived at 102 final codes with an inter-rater reliability of 96.13\% and a Cohen's kappa coefficient of 0.753. We then examined the coding categories, cross-referenced them with the data units, and analysed them for overarching themes, which we reviewed together. We identified seven overarching themes, and in this paper, we report on four themes that align with our research question and relate to the user experience of watching the body play games by sharing bodily control with EMS. These four overarching themes were developed using 76 of the 102 final codes, and these 76 codes comprised 658 of the 930 data units.

\section{FINDINGS: USER EXPERIENCE THEMES OF FUSED SPECTATORSHIP}\label{sec5:UXthemes}
In this section, we articulate and discuss four overarching user experience themes: 1) developing bodily relationships through loaning bodily control, 2) engaging with the ambiguity of a computer-controlled body, 3) customising spectatorship through active bodily involvement, and 4) appreciating the playfulness of the computer-controlled body.

\subsection{Theme 1: Developing Bodily Relationships Through Loaning Bodily Control}\label{sec5.1:theme1}
In this theme, we explore how participants' relationship with their body changed over time due to loaning bodily control, as well as their reflections on the experience of watching their body play involuntarily and how it influenced their perception of themselves. This theme comprises 168 data units, which have been grouped into 20 codes divided into two sub-themes: 1) surprising relationships with their computer-controlled body (91 data units), and 2) engaging with their altered self by loaning bodily control to a computer (77 data units).

\subsubsection{Surprising Relationships with their Computer-Controlled Body}\label{sec5.1.1:theme1}
Seven participants discussed how they developed surprising relationships with their computer-controlled body. They spoke about how it initially started as being (P1) \textit{``weird''} and \textit{``ecstatic''} (P4), then became (P10) \textit{``second nature''} over time. P4 elaborated on the ``ecstatic'' experience and said, \textit{``I knew I was not telling my brain to move my hand, yet it was moving.''} P6 called the EMS system ``someone'' and said, \textit{``It felt like `someone' was doing things to my body. I could not exactly comprehend this in my head.''} Participants also described the intelligence of the machine. P1 said, \textit{``While I knew there was no intelligence there, it was tough to believe it was not.''} Three participants also deliberately gave away complete control over their hand movements to the machine and let `it' play the games. P11 said, \textit{``I just let `it' do its thing.''} Five participants described how they had a collaborative relationship with the machine. P2 said, \textit{``All three agents, my hands, the system and my brain, worked together. Sometimes, it was easy to comprehend when they were collaborating (\'{I}dio game) and not so easy to comprehend when they were competing in the other two games.''} Three participants even contrasted this experience with watching others play digital games. P10 said, \textit{``This was a unique experience as I was not just comprehending my thoughts and emotions, but also my moving body.''} These results suggest that loaning bodily control to a computational EMS system can create a sense of surprise, excitement, and even collaboration with the machine.

\subsubsection{Engaging with their Altered Self by Loaning Bodily Control to a Computer}\label{sec5.1.2:theme1}
\begin{figure}[hbt!]
    \centering
    \includegraphics[width=1\textwidth]{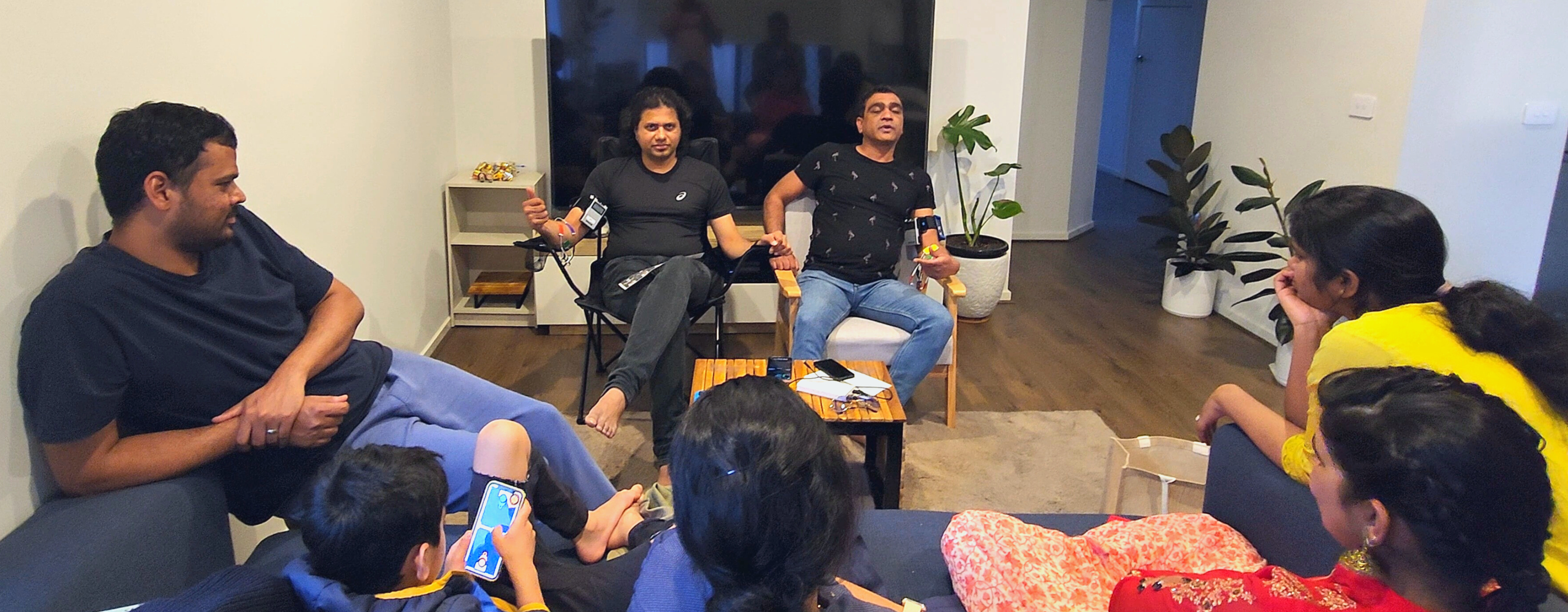}
    \caption{Shows two participants using the game kit socially. The two participants wear one EMS device to experience the games together. We can also see family members watching the participants' bodies move involuntarily.}
    \Description{Shows two participants using the game kit socially. The two participants wear one EMS device to experience the games together. We can also see family members watching the participants' bodies move involuntarily.}
    \label{fig:familywatching}
\end{figure}

Five participants described their experiences of becoming more aware of their agency over their body. The experience reminded P4 about how they felt when watching themselves in a video, for example, when they were dancing and said, \textit{``It makes me feel insecure when I watch myself do things as I notice all the small, weird things that I am not conscious of when I did the action originally.''} Comparing this to the study experience, P4 said, \textit{``I am not insecure, as it is not me who is doing the movements to my body; it's the machine.''} P7 and P8 felt like they were one with the machine and provided entertainment to their family members (Fig. \ref{fig:familywatching}). P8 said, \textit{``I was closing my eyes and guessing what the EMS hands were doing and letting my family members enjoy the game along with me.''} While P7 felt like an entertainer, P1 felt lazy: P1 described themselves as a \textit{``meat bag''} and explained, \textit{``The technology made me feel `this' way, and it surprises me how easily my body can be controlled, although I know I still have bodily ownership of the experience.''} Reflecting on the experience, P9 described that they experimented with various intensity levels to understand their endurance levels. They said, \textit{``I was cranking up the EMS intensity levels just to play around and understand my endurance levels.''}

These findings indicate that participants can assume different roles, some of which may lead to a feeling of loss of control or being controlled by the computer. Therefore, the act of loaning bodily control to technology could have psychological implications for participants' sense of agency. As such, to address this, designers could be mindful of how they design their studies and systems, focusing on promoting users' agency and ownership over their experiences, which we discuss in Section \ref{sec7:Discussion}.

\subsection{Theme 2: Engaging with the Ambiguity of a Computer-controlled Body}\label{sec5.2:theme2}
In this theme, we discuss participants' engagement with the ambiguity of a computer-controlled body. This theme describes 228 data units, which equates to 16 codes that have been divided into two sub-themes: 1) the impact of the control loop's timing on understanding the computer-controlled body (71 data units), 2) the influence of sound on engaging with and understanding the computer-controlled body (84 data units), and 3) decoding the meaning of computer-controlled movements to gain bodily knowledge of gameplay (73 data units).

\subsubsection{The Impact of the Control Loop's Timing on Understanding the Computer-controlled Body}\label{sec5.2.1:theme2}
Eleven participants discussed the impact of the time given for each part of the control loop on their ability to understand how the system controlled their body. Participants commented on the time provided to interpret the game's results before being confirmed on the screen. P5 said, \textit{``The time given to interpret the result was not a lot, especially for the Godai game, as the rules were so complex.''} P2 suggested that \textit{``if another 2-3 seconds were added, it would have been perfect.''} These results suggest that time affects participants' abilities to understand the gameplay, and if less time is given, it could make the gameplay ambiguous to the spectator.

Three participants also expressed a desire for onboarding to understand the timing of the control loop. P1 suggested, \textit{``Just like calibration before the experience of watching the games, I would have liked to learn how the rounds of gameplay work,''} while P7 recommended that \textit{``there should be an option to understand how the control loop works in the calibration screen itself.''} Four participants also expressed a desire to control the time parameter for each phase of the spectatorship. These results suggest that as participants could not control the time allocated to each part of the loop, it sometimes gave them less time to comprehend the meaning of what the EMS hands were doing, making their spectating experience ambiguous.

\subsubsection{The Influence of Sound on Engaging with and Understanding the Computer-controlled Body}\label{sec5.2.2:theme2}
While all participants tried using the three sound modes during the study, ten indicated they enjoyed the two-pitched sound feedback. One of the two remaining participants (P1) had a hearing aid and said, \textit{``The sound produced by the microcontroller devices was not loud enough even with the volume turned up.''} Participants described the sound feedback associated with the countdown feature, and P10 said, \textit{``It was dramatic''} and felt \textit{``game-like''}. P7 added to this narrative and said, \textit{``It made me learn the rhythm of this part of the control loop and get used to it.''} P11 also said, \textit{``Every time this sound was playing along with the visual of the 3-2-1 shown on the screen, I tapped my toes.''} These results suggest that clear audio feedback helps to reduce ambiguity, as it reinforces understanding and aids in learning the meaning associated with the bodily gestures made by the EMS-controlled hands while spectating the games.

\begin{figure}[hbt!]
    \centering
    \includegraphics[width=1\textwidth]{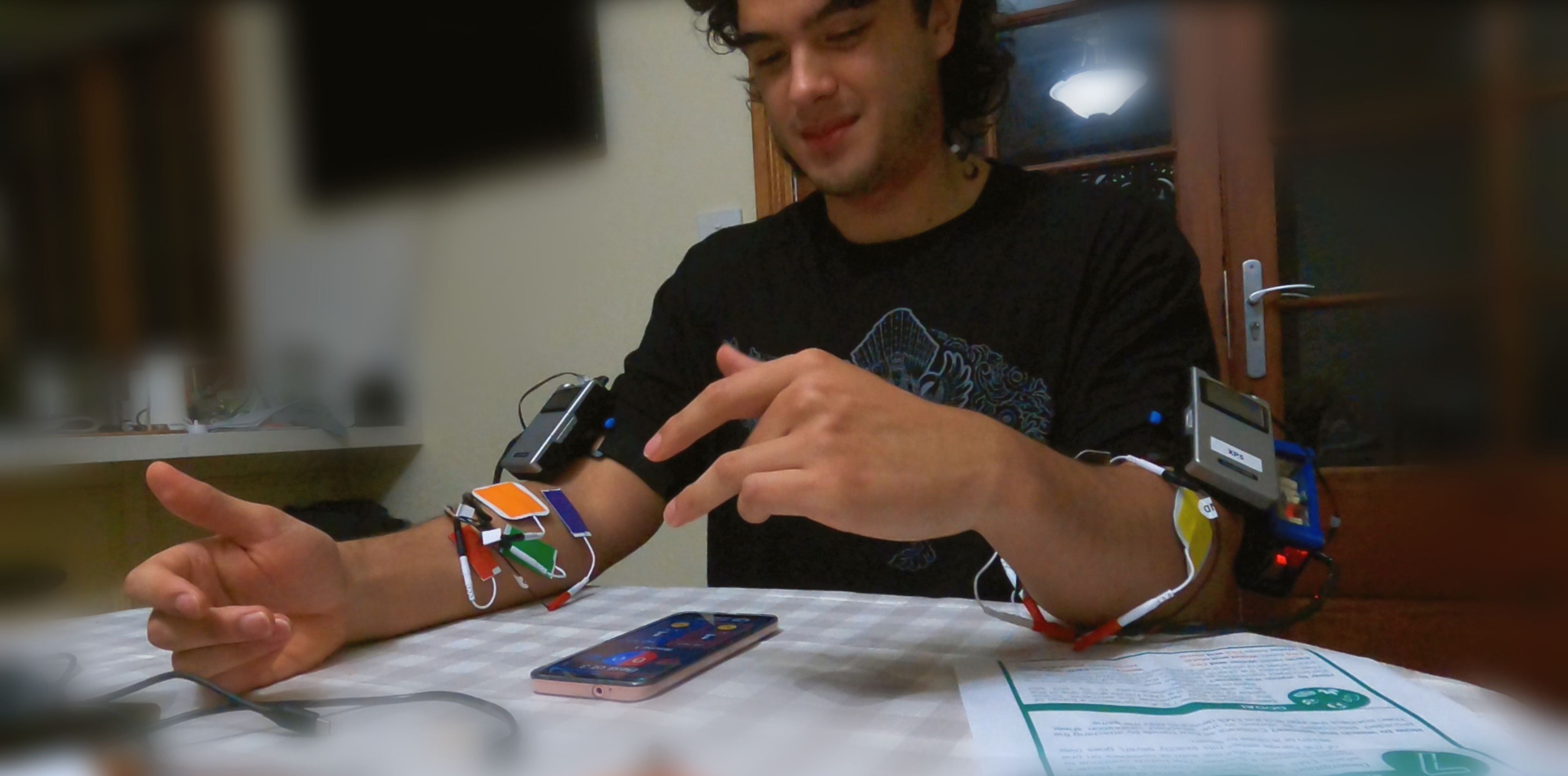}
    \caption{A participant looks at the phone to confirm their interpretation of the EMS-controlled hand's resulting gesture.}
    \Description{A participant looks at the phone to confirm their interpretation of the EMS-controlled hand's resulting gesture.}
    \label{fig:participantlookingatphone}
\end{figure}

Five participants described that, at times, they experienced all the games by closing their eyes. P8 said, \textit{``I liked that I could also rely on my ears to understand the hand gesture made by the EMS hands.''} These participants specifically described the open palm gesture, representing water in the Godai game and ``5'' in the Ept\'{a} game. P6 said, \textit{``When I close or open my eyes, I do not know when this gesture was performed. The sound was helpful during these times.''} P8 described how they did not focus on the sounds initially because of the learning curve for understanding the games. P12 also felt this way and said, \textit{``The sounds were confusing in the beginning. I had to focus on the sound to associate different gestures with different sounds.''} These results show that while sound feedback can reduce ambiguity, it could take time to understand what it means. Moreover, sound feedback could be helpful primarily when the EMS-controlled hands perform a gesture that does not actuate the body, reducing the ambiguity of such moves.

\subsubsection{Decoding Meaning of Computer-controlled Movements to Gain Bodily Knowledge of Gameplay}\label{sec5.2.3:theme2}
Nine participants described how the gameplay, which was sometimes ambiguous, made them gain bodily knowledge. They described how they appreciated that the results were hidden initially. For example, P8 added, \textit{``It made me intensely look at how my hand moved to try and interpret it without clicking on the reveal button on the application.''} Participants also described how they went back and forth between watching their hands and the phone screen to learn the gestures and their associated meanings for each game. P11 said, \textit{``I was new to EMS and the sensations it was causing to my body. I was getting used to these sensations and autonomous movements, and I was glad that I could click the reveal button to see the results and make associations.''} P9 added, \textit{``After the actuation, I always used to think about what the gesture meant in my head and then look at the screen for confirmation''} (Fig. \ref{fig:participantlookingatphone}).

These results suggest that participants paid close attention to the movements of their EMS-controlled hands and engaged in interpretation to create meaning, as the results on the screen were hidden by default. This ambiguity created a sense of curiosity and made them engage with their physical body as part of the learning experience.

\subsection{Theme 3: Customising Spectatorship through Active Bodily Involvement}\label{sec5.3:theme3}
In this theme, we discuss participants' experiences of customising their spectatorship through active bodily involvement. This theme describes 159 data units, which equates to 20 codes that have been divided into three sub-themes: 1) adaption of game order to manage cognitive load (56 data units), 2) the role of relaxation in loaning bodily control (42 data units), and 3) inward focus leads to self-exploration and bodily awareness (61 data units).

\subsubsection{Adaption of Game Order to Manage Cognitive Load}\label{sec5.3.1:theme3}
All participants described their experience of experimenting with the order of playing the games. Three participants with prior experience using EMS said they preferred playing competitive games first. In contrast, eight participants who did not have prior experience using EMS said they were more comfortable with playing the collaborative game of \'{I}dio first than the competitive games. P2 said, \textit{``Although I played Godai in the pre-study, I realised I should have played this collaborative game first while trying the system at home. It made me feel like I didn't have to do much and watch the hands make their gestures.''} P7 added to this narrative, saying, \textit{``There was very little cognitive load when playing \'{I}dio. I just had to watch my EMS hands match each other.''} All eight participants described the ultimate order in which they would like to experience these games in future. P10 said, \textit{``I would like to go with \'{I}dio first, then \'{E}pta and Godai. It felt like increasing the difficulty organically with this order.''} These results indicate that starting with collaborative games can be less demanding cognitively and might help participants become familiar with a computer controlling their bodies. However, this preference may not apply to participants with prior EMS experience, as they found it engaging to start with competitive games.

\subsubsection{The Role of Relaxation in Loaning Bodily Control}\label{sec5.3.2:theme3}
Ten participants reported their experience of relaxing their muscles when loaning bodily control to a computer. Out of these ten participants, three had prior EMS experience. P1 said, \textit{``I still needed to relax my body.''} They added, \textit{``I appreciated the breathing feature, which reminded me that I should relax and got used to the feeling of relaxing.''} P2 added to this topic, saying, \textit{``I was able to relax my muscles better by using this feature and let the EMS take control of my hand.''} Inexperienced EMS participants took a few goes before they understood why they should relax their body. For example, P6 said, \textit{``I realised the surprise my body felt after the first stimulation and that relaxing my body was important. This made me appreciate this breathing feature even further.''} Three participants skipped this breathing screen, and P2 said, \textit{``I am a very eager person. I wanted to experience the body play games. However, since I experienced EMS sensations, I knew when to relax my body to let the EMS control my hands.''}

\subsubsection{Inward Focus Leads to Self-exploration and Bodily Awareness}\label{sec5.3.3:theme3}
Ten participants discussed their experiences of focusing inward, leading them to explore their body and become aware of their hands. P2 said, \textit{``I always felt what the EMS was doing to my hands.''} Five participants also spoke about all the various bodily senses involved during this experience. P8 said, \textit{``Initially, I was only focusing on my body. However, as I got the hang of the games, I could also differentiate between the sounds. I was also engaged with the gameplay by watching my hands with my eyes.''}

Six participants spoke about EMS difficulties that made them focus intently on the actuations of the computer. P6 said, \textit{``The three-finger and middle-finger gestures were working extremely well for me. But the bigger hand movements, the inward and outward, felt very similar.''} P5 added that \textit{``I was calibrating them correctly. But I am not sure why they were feeling similar during gameplay. I had to focus intently on my hands to differentiate between these two movements.''} P11 described these two body movements and said, \textit{``I could feel my skin and still understand what was happening even when the EMS was fully moving my hand.''}

There appeared to be a difference regarding looking inward between participants with and without prior EMS experience. Three participants with previous EMS experience expressed how they were letting go of their body and being relaxed even while knowing what the EMS was doing. P1 said, \textit{``It was relaxing, and I felt like I was focused mindfully on the experience.''} Participants who were using EMS for the first time or using EMS for the first time in the context of play said, (P10) \textit{``I just wanted to explore my body's capabilities.''} These results also suggest that prior experience with EMS can impact how participants experience the fused spectatorship, with some participants being able to relax and let go of their body. In contrast, others were more focused on exploring their body's capabilities.

\subsection{Theme 4: Appreciating the Playfulness of the Computer-controlled Body}\label{sec5.4:theme4}
In this theme, we discuss participants' experiences of appreciating the playfulness of the computer-controlled body. This theme describes 103 data units, which equates to 12 codes that have been divided into three sub-themes: 1) anticipating and predicting game outcomes through bodily sensations (57 data units) and 2) competitiveness due to laterality (46 data units).

\subsubsection{Anticipating and Predicting Game Outcomes through Bodily Sensations}\label{sec5.4.1:theme4}
Seven participants described how they tried to understand what was happening in the game and how they tried to anticipate the results before their hands visually moved to show the gestures. P10 said, \textit{``Anticipating and predicting what was happening was the fun part of this experience.''} P2, who also compared this form of anticipation to anticipating when spectating about sports in general, said, \textit{``I want to predict what will happen. Sometimes, when I am watching a sport and if I know the player well, I will predict better than if I do not know the player well.''} P2 contrasted spectating games with their hands and said, \textit{``It was different to spectating games outside the body. I have more data points.''} P3 added to this narrative and said, \textit{``As this was happening on my body, I could feel the sensations even before the result was shown on the body. Therefore, I tried to make predictions.''} P7, who also tried to make predictions, said, \textit{``Sometimes the predictions were correct, but most of the time did not match what I was thinking in my head.''} 

Three participants closed their eyes to sense their body playing games. Upon questioning, P12 said, \textit{``I was more focused, and I could feel the hand and make the association with the sound playing synchronously. This helped me make better predictions and better engage with watching the games.''} Other participants who did not close their eyes said, \textit{``I did not close my eyes as the phone screen was not revealing the results by default anyways. So, it was as good as closing my eyes. I could not cheat unless I deliberately clicked the reveal button''} (P11). These findings suggest that bodily sensations can be a key factor in the anticipation and prediction of game outcomes, adding to the immersive experience of fused spectatorship.

\subsubsection{Competitiveness Due to Laterality}\label{sec5.4.2:theme4}
Ten participants discussed a sense of competitiveness that they felt, while the remaining two did not care about winning or losing. P1 said, \textit{``It was my own body, and I did not care which hand won or lost.''} P2, who also experienced the games using one of their hands and another person's hand, said, \textit{``While playing alone, I didn't feel competitive at all. I thought I would be competitive. But when I watched the games on my body with another person, I wanted my hand to win, even though I knew I was not making the gestures.''}

Laterality also seemed to play an unconscious role \cite{Corballis_2012}. Two participants only realised they were biased toward one of their hands after reflecting on their experience during the post-study interview. P9 said, \textit{``It took me a while to learn the gameplay and understand what was happening to my body. I never realised until now that I was biased toward my right hand. I remember trying to think that my right hand should win!''} In contrast, two participants described how they just were amazed and enjoyed their body moving autonomously. P10 said, \textit{``It was just amazing to see my hands move. I was not focusing on the gameplay, winning, or even trying to learn the rules.''} P7 said, \textit{``There was competition at the beginning when I had to learn, but it dumbed down after I finished playing all the games and started feeling repetitive. It just became a mindless activity, just like watching TV.''} These results could possibly suggest that participants experienced a sense of competition even if it was their own body.

In the next section, we reflect on these themes and discuss the experiences that transpired because of \textit{Fused Spectatorship}.

\section{FUSED SPECTATOR TYPES AND THE ASSOCIATED BEHAVIOUR}\label{sec6:fusedspectatortypes}
Prior work has categorised player types, immensely benefiting the play community \cite{Bartle_1996}. Following suit, previous research on digital game spectators identified five types relevant to spectating digital games on YouTube over a screen \cite{Golob_Krasevec_Crnic_2021}. We found that the participants' behaviours did not align well with these existing types, possibly due to the more bodily character of our spectating experience. Therefore, to complement the prior work \cite{Golob_Krasevec_Crnic_2021}, we now present the spectator types that emerged from our study and then compare them in section \ref{sec7:Discussion}.

\begin{figure}[hbt!]
    \centering
    \includegraphics[width=1\textwidth]{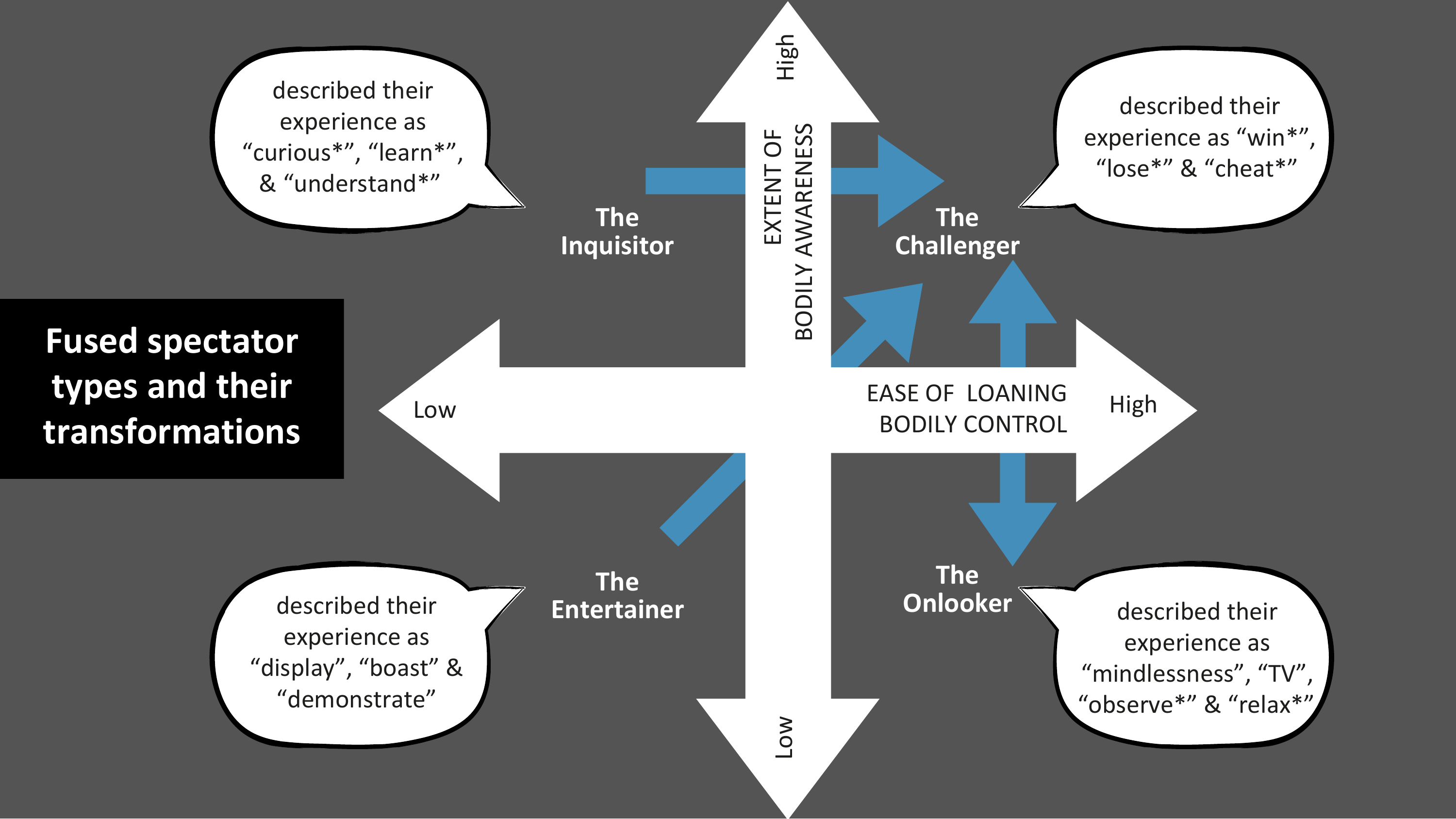}
    \caption{Our dimensional diagram situates the Fused Spectator types on four quadrants, describing participants' experiences of \textit{Fused Spectatorship}. Three spectator types (The Inquisitor, The Entertainer and The Onlooker) can transform into another spectator type (Challenger), with ``The Challenger'' also being able to transform into ``The Onlooker''.}
    \Description{Our dimensional diagram situates the Fused Spectator types on four quadrants, describing participants' experiences of \textit{Fused Spectatorship}. Three spectator types (The Inquisitor, The Entertainer and The Onlooker) can transform into another spectator type (Challenger), with ``The Challenger'' also being able to transform into ``The Onlooker''.}
    \label{fig:UXspectators}
\end{figure}

\begin{figure}[hbt!]
    \centering
    \includegraphics[width=1\textwidth]{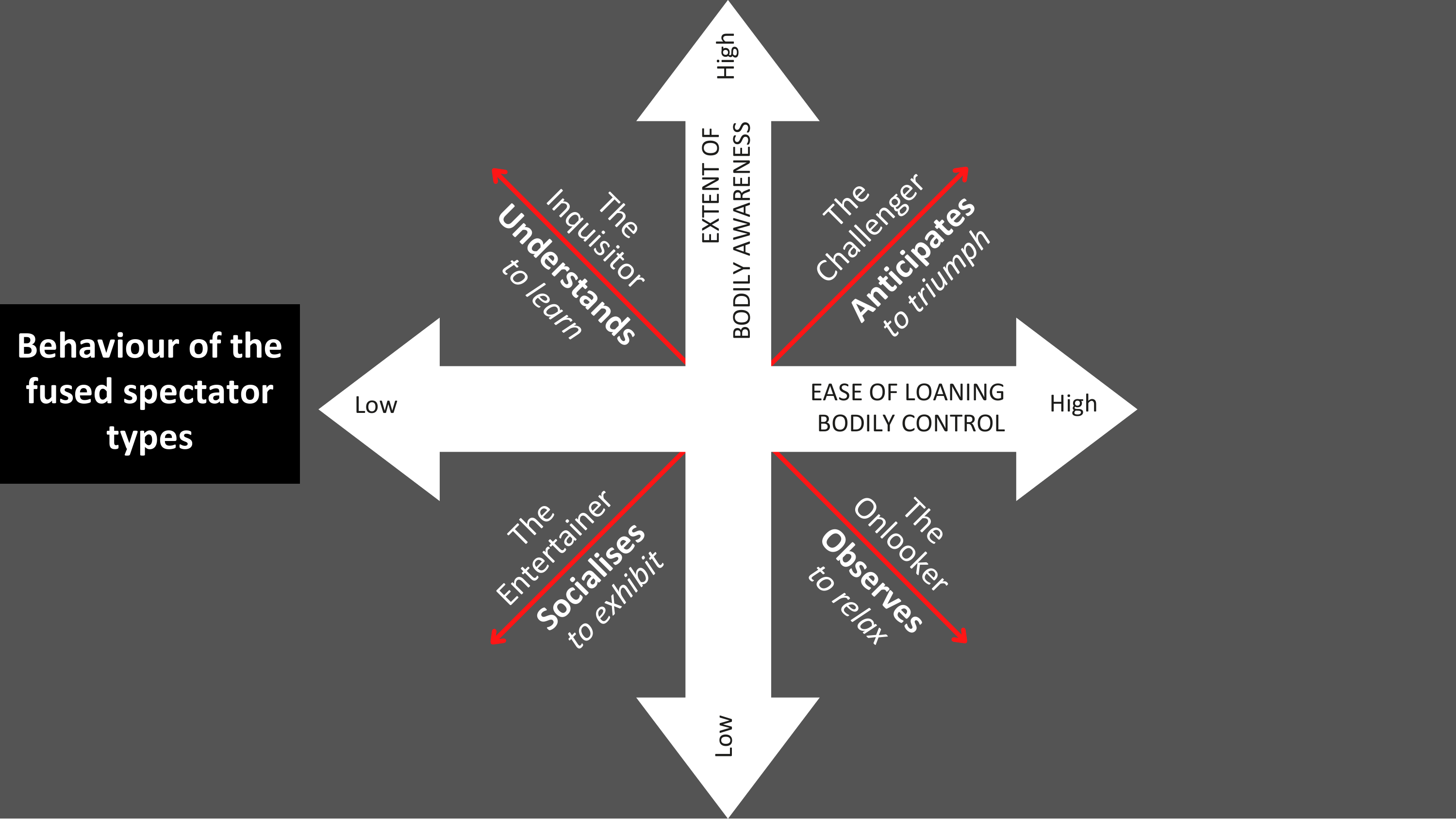}
    \caption{The associated behaviour of the four fused spectator types.}
    \Description{The associated behaviour of the four fused spectator types.}
    \label{fig:specbehaviours}
\end{figure}

All our study participants exhibited varied behaviours, influenced by their experience with EMS, willingness to engage with our EMS system, and their understanding of the game rules. These behaviours led us to identify four fused spectator types: The Inquisitor, The Challenger, The Onlooker, and The Entertainer. These spectator types are situated on four dimensions plotted on a dimensional diagram, with the ``cause'' on the x-axis and the ``effect'' on the y-axis representing the ``ease of loaning bodily control'' and ``extent of bodily awareness'', respectively (Fig. \ref{fig:UXspectators}). We use these dimensions as our study found that loaning bodily control was easy for some participants but difficult for others, which we used to categorise their experiences into four spectator types plotted on a dimensional diagram. We also analysed how their experiences transformed over time, represented by blue arrows in Fig. \ref{fig:UXspectators}. Each type exhibited a generalised behaviour (Fig. \ref{fig:specbehaviours}), and we offer one design consideration to support each behaviour and create engaging future fused spectatorship experiences.

\subsection{The Inquisitor Understands to Learn}\label{sec6.1:inquisitorspectatortype}
Our study identified a particular type of spectator - The Inquisitor - who enjoyed learning about the game rules and their body and was consequently willing to undergo the initial discomfort associated with no longer having complete control over their hands (Themes 1, 3 and 4). This type included one participant with, and six without prior EMS experience, demonstrating a low level of ease in loaning bodily control and a high level of bodily awareness. Notably, two Inquisitors transitioned into The Challenger type over time as they started to care about competing, demonstrating how participants' behaviours and experiences could evolve.

The Inquisitor type is characterised by a desire to learn and use strategies such as closing their eyes to perceive the EMS actuated gestures. The Inquisitor prefers tactile feedback over sound feedback. Furthermore, our study revealed that Inquisitors value the breathing feature. This highlights the importance of matching feedback mechanisms with participants' preferences and designing features for relaxation.

\subsubsection{Design Consideration for the Inquisitor: Facilitate relaxation to help ease sharing bodily control}\label{sec6.1.1:inquisitorspectatortype}
The Inquisitor values learning about and comprehending the system's capabilities, which is crucial for their engagement. As prior research suggests, introducing new entertainment mediums with an onboarding process is crucial \cite{Reeves_Benford_OMalley_Fraser_2005}, particularly when expecting someone to loan bodily control to a computer \cite{Benford_Ramchurn_Marshall_Wilson_Pike_Martindale_2021}. Interestingly, our study found that the Inquisitor type, mainly comprising participants without EMS experience, found the breathing feature helpful during their onboarding process. Thus, to aid the onboarding process for the Inquisitor type, we suggest extending the concept to the design of \textit{Fused Spectatorship}. This can be achieved by incorporating features that help relax participants' muscles and to ease the process of loaning bodily control.

\subsection{The Challenger Anticipates to Triumph}\label{sec6.2:challengerspectatortype}
The Challenger spectator type anticipates to triumph. This spectator type was motivated to succeed and enjoyed learning the rules to achieve victory. Four participants demonstrated the characteristics of this type, with high ``ease of loaning bodily control'' due to their prior EMS experience (Themes 2 and 3). However, they also exhibited a high ``extent of bodily awareness''. They were highly driven to win and even considered cheating to ensure their dominant hand emerged victorious.

The behaviour of this spectator type is characterised by anticipation of success. These participants were primarily focused on anticipating the ambiguous EMS actuations and generating a feeling of winning. They relied on specific strategies such as closing their eyes to anticipate the EMS actuations or listening to the two-pitched sound feedback to achieve this. The "reveal button" feature was particularly appreciated, adding an element of challenge by hiding the results. We also observed that they would mentally predict and reveal the outcome on the screen to check their anticipation, enhancing their overall engagement.

\subsubsection{Design Consideration for the Challenger: Harness the computer's ambiguity to foster anticipation}\label{sec6.2.1:challengerspectatortype}
To support the Challenger spectator type's desire to anticipate and achieve a feeling of "winning", it can help them anticipate the ambiguous movements of a computer-controlled body. Hiding the game's result on the screen by default allows them to anticipate what the computer-controlled hands might do without seeing the result, and then reveal it after checking if they were correct. Prior work has suggested that EMS actuations are usually understandable, even if sometimes ambiguous, because any change in the muscle's orientation can lead to an incomplete actuation \cite{Jonell_Lopes_2016}. Moreover, ambiguity is considered a valuable design resource as it can create a sense of mystery and surprise, which benefits games \cite{Gaver_Beaver_Benford_2003,Sutton-Smith_2001}. Therefore, if designing for the Challenger, we suggest designers consider hiding any digital confirmation of actuation output to retain ambiguity to foster anticipation.

\subsection{The Onlooker Observes to Relax}\label{sec6.3:onlookerspectatortype}
The Onlooker type does not have much interest in learning about their body or the game but instead finds pleasure in surrendering bodily control for a relaxing experience. This type mainly comprised participants with prior EMS experience who enjoyed the EMS actuation (Themes 1 and 4). They were more interested in watching their hands move involuntarily without caring much about the meaning of those movements.

The behaviour of the Onlooker type is characterised by a desire to observe and relax. While they had no trouble loaning control over their body movements to the EMS, they were not actively aware of their movements. The sound of the EMS-controlled hands was the primary source of engagement for this type of spectator, indicating their disinterest in actively engaging with the game. Interestingly, one participant of this spectator type demonstrated qualities associated with the Challenger type, i.e., they chose to be involved or uninvolved in the act of spectatorship, like traditional spectator experiences.

\subsubsection{Design Consideration for the Onlooker: Subconsciously support awareness of control exchange using subtle sounds}\label{sec6.3.1:onlookerspectatortype}
To support the Onlooker, it can be beneficial to support awareness of control exchange using subtle sounds. Our study showed that they enjoyed the first part of the two-pitched sound feedback to be aware of when the system was about to control their body. Previous research has already emphasised the need to provide feedback to inform users when a machine and a human are about to exchange control over each other \cite{Benford_Ramchurn_Marshall_Wilson_Pike_Martindale_2021}. To extend this notion to \textit{Fused Spectatorship}, we suggest using subtle sounds to support the Onlooker in becoming aware when the system is about to take control, which might support their relaxation experience.

\subsection{The Entertainer Socialises to Exhibit}\label{sec6.4:entertainerspectatortype}
This spectator type, the Entertainer, focused on showcasing their involuntary movements to others. Themes 1 and 2 revealed that they engaged not to learn like The Onlooker, but rather to exhibit their relationship with the computer to friends and family. The Entertainer participants could easily relinquish control over their body movements to the computer; however, they were not actively aware of their movements. They enjoyed displaying the visuals of the EMS-controlled hands moving and showcasing their involuntary actions to others. Interestingly, we observed that some participants transitioned to ``The Challenger'' type over time, indicating a potential shift from pure entertainment to valuing the experiences associated with learning and winning.

Our study suggested that the Entertainer participants often endured the uncomfortable sensations associated with EMS, experimenting with the EMS intensity to entertain others and showcase their endurance level. Additionally, they utilised the modularity of the game kit and rules to engage socially with our games, which were initially designed for one person. We were surprised that the Entertainers perceived the "comprehensiveness" of our EMS system as ``modularity'', which was not our conscious design choice. We note that this type wanted to exhibit their body as a "toy" and relied on their social counterparts to interpret the outcomes of the computer controlling their body.

\subsubsection{Design Consideration for the Entertainer: Enable social experiences through modular design}\label{sec6.4.1:entertainerspectatortype}
To support ``The Entertainer'', it might be beneficial to support a social aspect rather than solely focusing on the spectating of the game itself. Our study suggested that this spectator type enjoyed the modularity of our system as it allowed for social experimentation. They liked sharing the system's modular hardware with another spectator for a shared experience. Prior work already highlighted the importance of modularity when designing systems for playful interactions to foster flexibility and creative exploration \cite{Lund_Marti_2009}. We suggest designers extend this notion of modularity to the design of fused spectatorship experiences by creating modular systems to support The Entertainer in engaging socially. This approach could enable them to show off socially by exploring their bodily limitations, which prior work suggested can be key for entertainment \cite{Isbister_Mueller_2015}.

\section{DISCUSSION}\label{sec7:Discussion}
We now reflect on our results in sections \ref{sec5:UXthemes} and \ref{sec6:fusedspectatortypes} to compare spectator types (section \ref{sec7.1:Discussion - ComparingSpecTypes}) and discuss whether our participants were indeed spectating or playing (section \ref{sec7.2:Discussion - PlayingOrSpectating}). We also discuss the future of our \textit{Fused Spectatorship} approach (section \ref{sec7.3:Discussion - FutureofFS}) and its ethical design (section \ref{sec7.4:Discussion - Ethics}).

\subsection{Comparing Spectator Types: Overlapping characteristics and motivations}\label{sec7.1:Discussion - ComparingSpecTypes}
\begin{figure}[hbt!]
    \centering
    \includegraphics[width=1\textwidth]{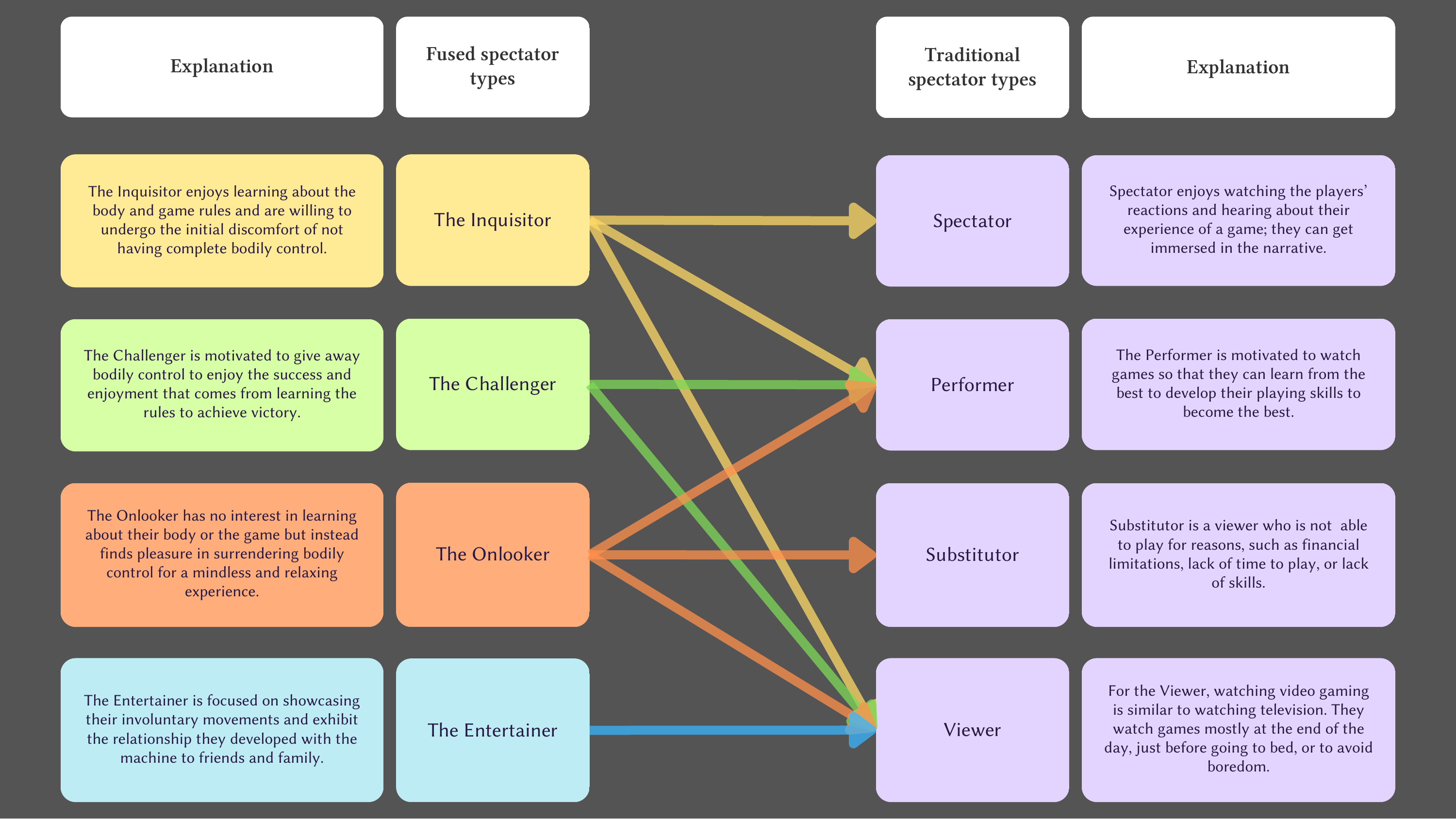}
    \caption{Comparing fused and traditional \cite{Golob_Krasevec_Crnic_2021} spectator types.}
    \Description{Comparing fused and traditional spectator types.}
    \label{fig:comparingspectypes}
\end{figure}

We now discuss the overlapping characteristics and motivations between four of the five traditional spectator types identified by Golob et al. \cite{Golob_Krasevec_Crnic_2021} (Spectator, Performer, Substitutor, and Viewer) and our fused spectator types. The Selector type from Golob et al.'s \cite{Golob_Krasevec_Crnic_2021} is driven by pragmatic considerations and cost-benefit analyses of games. This type does not align well within our study, as our participants were not intending to buy EMS games but rather to experience a novel system. While comparing these spectator types, we believe it is important to note that the relationships between the traditional and \textit{Fused Spectatorship} are not one-to-one or uniform; instead, they intersect in different ways, reflecting the complexity of spectating and play (Fig. \ref{fig:comparingspectypes}).

The Inquisitor, who enjoys learning about the body and the game rules, may share some similarities with the Spectator, Performer, and Viewer \cite{Golob_Krasevec_Crnic_2021}. The Spectator, drawn to the player's narrative when watching streams on YouTube, shares behavioural aspects with the Inquisitor. Particularly, the Inquisitor is also curious to understand and build a relationship with the system through which they experience the gameplay on their body. The Performer's focus on learning and improving gameplay can also overlap with the Inquisitor's interest in understanding the sensations of control and interaction on a bodily level.

The Challenger, who seeks an active role in enhancing their gameplay experience by letting the system take bodily control, shares characteristics with the Performer and Viewer. Both the Challenger and Performer are interested in improving their gameplay, albeit with different methods (observation for the Performer and direct bodily involvement for the Challenger). The Viewer's motivation for entertainment can also overlap with the Challenger's desire for a more engaging and interactive gameplay experience - even though they know they are not playing the game with their body.

The Onlooker, motivated by the desire to explore the possibilities and potential of their body, shares characteristics with the Substitutor, Viewer and Performer. The Onlooker's interest in exploring new experiences for relaxation and watching the body move on its own speaks to the Substitutor seeking gameplay as a replacement for playing themselves. Additionally, the Viewer's pursuit of personal entertainment and relaxation can align with the Onlooker's curiosity in exploring the fused spectatorship experience. As we observed in section \ref{sec6.3:onlookerspectatortype}, the Onlooker can sometimes snap out of the relaxation mode to actively watch the gameplay, just like the performer - speaking to the complexity of comparing these spectator types.

Lastly, The Entertainer, who actively engages in social interactions and shared bodily performances through the system, shares overlapping characteristics with the Viewer. The Entertainer's focus on entertaining others while not being in control of their actions is especially unique to \textit{Fused Spectatorship}. It can intersect with The Viewer's motivation for entertainment, albeit it is personal entertainment that this spectator type seeks, unlike the Entertainer, whose desire is to create collective experiences and forge connections with others.

\subsection{Fused Spectatorship: Spectating or playing?}\label{sec7.2:Discussion - PlayingOrSpectating}
A surprising result from our study was that 11 participants could not distinguish between their roles - were they spectating or playing? This observation resonates with earlier work, where participants were uncertain about the dichotomy between the ownership and control of the bodily actuations caused by EMS \cite{Kasahara_Nishida_Lopes_2019,Kasahara_Takada_Nishida_Shibata_Shimojo_Lopes_2021}. This is possibly because the actuation can make users feel like their control is being taken away while they still retain the feeling of bodily ownership \cite{Kasahara_Nishida_Lopes_2019,Kasahara_Takada_Nishida_Shibata_Shimojo_Lopes_2021}. However, this discovery of participants' affective experience caused by \textit{Fused Spectatorship} is interesting because it challenges a common definition of ``play'', which states that ``play can occur only if the player is in control'' \cite{Caillois_2001,Huizinga_2016}.

Traditional spectatorship enables a certain freedom - spectators can choose to engage or disengage themselves from the experience \cite{Golob_Krasevec_Crnic_2021}. However, in \textit{Fused Spectatorship}, the constant possibility of the computer taking over bodily control, even during non-active periods, maintains a thread of continuous engagement, resulting in a possibly constant immersion in the game experience. This persistent computer's ability to take over bodily control may have prompted caused our participants to struggle to differentiate between playing and spectating. Thus, \textit{Fused Spectatorship} intertwines the identities of spectator and player and introduces a fluid spectrum of engagement, challenging the binary norms. This feeling of ``owning the body'' - as a spectator, and ``controlling the body'' - as a player, blurs the spectator-player relationship, so much so that it may become hard to distinguish between them. This discovery means that play could be facilitated in scenarios where the spectator cannot control their body movements. 

\begin{itemize}
    \item For example, our approach could be helpful in motor rehabilitation practices. Medical professionals usually control the patient's movements using EMS, making them feel like they are just a spectator of their moving body \cite{Parry_Berney_Granger_Koopman_El-Ansary_Denehy_2013}. In such cases, our work could be useful by creating a more enjoyable experience and empowering patients to take ownership of their body movements. This could potentially improve their motivation and engagement with the rehabilitation process, leading to better outcomes. We encourage further research to explore the potential benefits of \textit{Fused Spectatorship} in other areas of healthcare and well-being.
    \item The potential use of \textit{Fused Spectatorship} in making gaming more inclusive such as for people with Ataxia, is another scenario where our work could be helpful. Ataxia \cite{Ashizawa_Xia_2016}, which involves poor muscle control, can make gaming difficult for those affected. However, using our approach, such people could feel like they have greater control over their bodily movements, allowing them to engage in movement-based experiences, ultimately contributing towards creating a more inclusive society.
    \item Another way our approach could be helpful is to transfer skills between a streamer and their spectators. In particular, we believe this could make the rather passive relationship between onlookers and players \cite{Knibbe_Freire_Koelle_Strohmeier_2021} more active. For example, this could be done by using Electromyography (EMG) sensors on the players and using EMS on the spectators to transfer skills between players.
\end{itemize}

In addition to these examples, our approach could lead to new ways of spectating and playing. Notably, some example ways in which game designers could explore using our approach in the realm of affective gaming \cite{Gilleade_Dix_Allanson_2005,Robinson_Wiley_Rezaeivahdati_Klarkowski_Mandryk_2020} are:
\begin{itemize}
    \item \textbf{Emotional mirroring:} Game designers could use our approach to enable emotional mirroring, where the spectators' movements, controlled by EMS, reflect the emotions or actions of characters in a game. For instance, in the context of an emotional scene within a narrative-driven game like ``Life is Strange'' \cite{de2018lifeisstrange}, designers could synchronise the protagonist's emotional state to move the spectator's body physically. The physicality of such spectating experiences could lead spectators to have a deeper emotional connection with the game, enhancing their overall experience \cite{Brown_Cairns_2004}.
    \item \textbf{Shared emotional experiences:} Our approach could also enable designers to create shared emotional experiences among spectators. For example, in a game like ``Journey'' \cite{TGC_2017}, where multiple players can share their journey across the internet, applying our approach could allow these players to share their emotional experiences. This shared experience could foster a stronger sense of connection, enhancing the social aspect of gaming \cite{Isbister_2010}.
\end{itemize}

While we believe that the potential of our approach is exciting, we now reflect on both technical (such as making the calibration process more manageable \cite{Knibbe_Freire_Koelle_Strohmeier_2021}) and ethical challenges \cite{Lopes_Chuang_Maes_2021} (such as how designers and users could engage with systems with which they can loan bodily control) in section \ref{sec7.4:Discussion - Ethics}.

\subsection{The Future of Fused Spectatorship}\label{sec7.3:Discussion - FutureofFS}
In this section, we situate the games we designed along with the examples from mass media and games outlined in our related work sections \ref{sec2.1:specagency} and \ref{sec:engagingthebodyspec} on a continuum (section \ref{sec7.3.1:Discussion - SpectatorAgency}). We then examine Fused Spectatorship's inherent qualities beyond the dichotomy between playing and spectating (Section \ref{sec7.3.2:Discussion - InherentQualFS}).

\subsubsection{Spectator Agency: From traditional to Fused Spectatorship experiences}\label{sec7.3.1:Discussion - SpectatorAgency}
\begin{figure}[hbt!]
    \centering
    \includegraphics[width=1\textwidth]{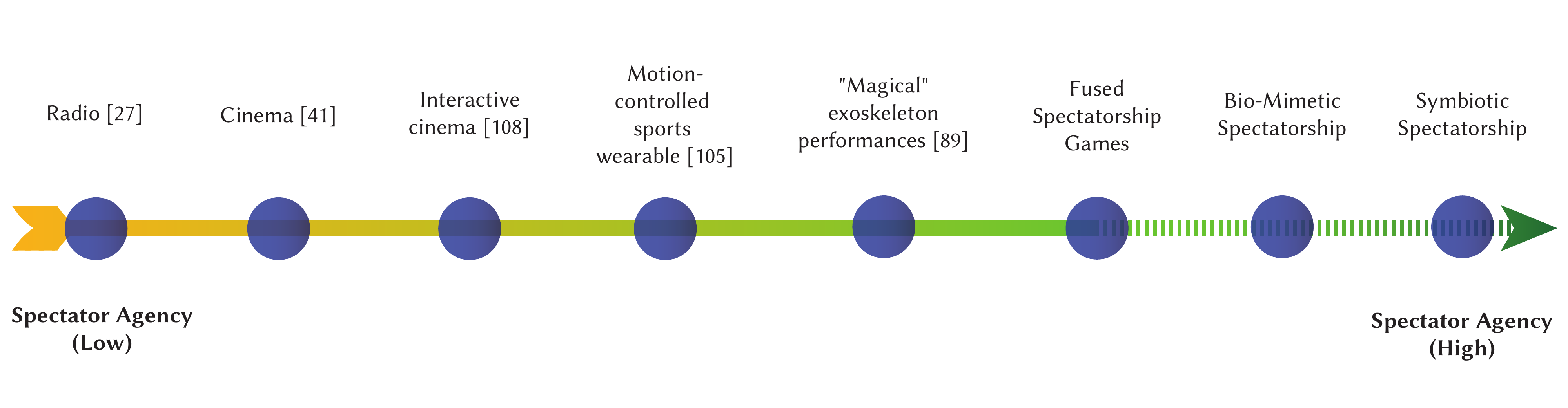}
    \caption{A continuum of the evolving spectator agency.}
    \Description{A continuum of the evolving spectator agency.}
    \label{fig:specagencyevolution}
\end{figure}

This section examines how the spectator's agency is evolving using the examples presented in sections \ref{sec2.1:specagency} and \ref{sec:engagingthebodyspec}, leading up to our \textit{Fused Spectatorship} games. We represent these examples along a continuum (Fig. \ref{fig:specagencyevolution}), showcasing the progression of spectator agency and the increasing levels of bodily involvement in spectating experiences. We also propose extending this continuum by introducing the ``Bio-Mimetic Spectatorship'' concept and advancing it to the ``Symbiotic Spectatorship'' approach.

At one end of the continuum, we have traditional spectating experiences, such as the radio \cite{Douglas_2004} and cinema \cite{Gunning_2004}, which offer little to no agency to the spectators, with the creators dictating the narrative. Moving along the continuum, we find more interactive experiences, such as interactive cinema \cite{Elnahla_2020_blackmirror} and motion-controlled wearables for sports spectating \cite{Tomitsch_Aigner_Grechenig_2007}. These examples allow spectators to have some agency over their experience, albeit still limited compared to the players or performers.

The ``magical'' performances involving exoskeletons, where an exoskeleton ``makes'' the wearer dance to a DJ performance \cite{Reeves_Benford_OMalley_Fraser_2005}, and our three games have a common element. In both these cases, the technology controls the wearer's body. However, in the case of the exoskeleton performances, \cite{Reeves_Benford_OMalley_Fraser_2005}, the technology controls the performer, and any spectators are simply spectating the involuntary movements. In contrast, in our work, the participant spectates an EMS system play the games by using their body. Our games represent the other end of the continuum, where spectators experience more agency than previously mentioned spectating experiences.

Moving further along the continuum, we introduce the ``Bio-Mimetic Spectatorship'' concept. In this advanced form of \textit{Fused Spectatorship}, devices such as EMG sensors or motion-tracking could capture a player's movements while playing movement-based games, like rock-paper-scissors. Designers could then actively engage the spectator's body by translating the player's movements into bodily movements that the spectator can experience using technologies like EMS. This way, the spectator's experience is more directly driven by another player's actions, rather than a computer controlling their body (like in our case). With this concept, there is possibly a greater extent of integration between spectators and players, and spectators experience a higher level of agency and bodily involvement in the gameplay.

Taking this idea a step further, we propose the ``Symbiotic Spectatorship'' approach, where the relationship between the player and the spectator is bidirectional and mutually influential. This would allow both parties to actively impact each other's gameplay experience. For example, the spectator could provide guidance and assistance to the player through their own bodily movements, which could be captured using wearable devices or motion sensors and translated into in-game actions. This approach could foster a deeply connected and immersive experience for both the player and the spectator, encouraging collaboration, social interaction, and a heightened sense of presence and agency, where the spectator and player distinction is blurred to a large extent.

The next section discusses the inherent qualities that lend themselves to the \textit{Fused Spectatorship} approach.

\subsubsection{Inherent Qualities of Fused Spectatorship}\label{sec7.3.2:Discussion - InherentQualFS}
The term ``fused'' refers to a unique mode of engagement that blurs the lines between playing and spectating by incorporating elements of both experiences. This fusion is achieved through the combination of technology, such as an EMS system in our case, and the spectator's physical involvement in the gameplay - transcending the traditional boundaries of play \cite{Seering_Savage_Eagle_Churchin_Moeller_Bigham_Hammer_2017}. In this section, we discuss the inherent qualities of our approach: physicality, social and inclusive.

\begin{itemize}
    \item \textbf{Physicality:} One of the inherent qualities of the \textit{Fused Spectatorship} approach is the emphasis on embodiment and, particularly, physicality. Unlike conventional game spectating, where physical engagement is often limited to clapping or jumping in joy, \textit{Fused Spectatorship} allows participants to experience the game through their bodies. We believe that this physicality can create an immersive and intimate connection between the spectator, the player and gameplay actions (Fig. 1), creating a heightened sense of presence and agency within the magic circle, the virtual space where the rules of play apply \cite{Huizinga_2016}.
    \item \textbf{Social interaction:} Another inherent quality of our approach we observed through our study is the potential for facilitating social interaction and shared experiences. This shared involvement in spectating gameplay facilitates social bonding and collaboration, fostering a sense of community among participants and enhancing the overall gaming experience \cite{Bekker_Sturm_Barakova_2009,Garvey_1974}. Besides the social quality of our approach, we also observed that the participants could spectate the games by closing their eyes.
    \item \textbf{Inclusive:} We also observed a heightened sensory engagement experienced by the participants in our study, which highlights another quality of our approach. By allowing spectators to close their eyes while spectating the games, our approach affords spectators to have a stronger reliance on proprioceptive and haptic sensations, which can lead to a deeper emotional connection with the gameplay \cite{Gatti_Pittera_Berna_Moya_Obrist_2017}. This rather unusual (especially compared to traditional digital gameplay) sensory engagement appears to augment the spectating experience and offers novel opportunities for inclusive game spectating, potentially catering to individuals with visual impairments or those who may benefit from a more tactile-focused experience of gameplay \cite{Baker_Ramos_Turner_2018}.
\end{itemize}

We acknowledge that these qualities of \textit{Fused Spectatorship} are not exhaustive, particularly because we focus on using EMS. We believe the potential for other technologies to achieve similar effects is worth considering. For example, future designers could consider including virtual reality (VR) or augmented reality (AR) when designing using our approach. These technologies can cut out the outside world to create immersive environments and provide spectators with a first-person perspective of the game, allowing them to feel like they are part of the action. Combining these technologies with other sensory inputs, such as haptics, has already enhanced player experience \cite{Lopes_Baudisch_2013}. Similarly, such systems could be used to explore and understand the effect on game spectatorship, which might benefit the community by gaining a deeper understanding of our approach.

Brain-computer interfaces (BCIs) also offer the potential to enhance fused spectatorship experiences. BCIs can be used to interpret the player's brain activity, allowing players, for example, to potentially more directly communicate emotions between themselves and the spectators. Conversely, this technology could enable spectators to control aspects of the game or influence the player's actions, adding a new layer of interaction and agency to the spectating experience \cite{Fang_Semertzidis_Scary_Wang_Andres_Zambetta_Mueller_2021}.

In summary, our approach offers a transformative approach to game spectatorship, emphasising physicality, social connections, and inclusivity. By exploring various technologies in the future, this approach might act as a stepping-stone to reshape and enrich the spectating experience, fostering a deeper bond between the spectator, player, and gameplay.

\subsection{Fused Spectatorship: Ethical design considerations}\label{sec7.4:Discussion - Ethics}
In this section, we discuss the ethical design considerations surrounding \textit{Fused Spectatorship} by examining two distinct yet interrelated aspects. First, we discuss the ethical considerations involved in the design and execution of our study, highlighting the measures taken to ensure participant safety and well-being and respecting their autonomy. Second, we delve deeper into the ethical design considerations that designers can use to mitigate potential harm when participants are asked to loan bodily autonomy to a computer.

\subsubsection{Ethical Considerations in the Design and Execution of our Study}\label{sec7.4:Discussion - StudyEthics}
Our study's results indicated that loaning bodily control to technology could have psychological implications for users' sense of agency and control (Theme 1). This section discusses our findings concerning our system and study design using three aspects of van de Poel's ethical framework \cite{van_de_Poel_2016}. While we use these three aspects, we acknowledge that others exist and might be useful to examine in future work. Through our work, we aim to extend this prior philosophical discourse by providing practical guidance for designers interested in creating ethical Fused Spectator experiences.

\begin{itemize}
    \item \textbf{Non-maleficence} is the first aspect that asks designers to question: what precautions are taken against possible risk? We took the following precautions to ensure the safety and well-being of our participants during the study. In addition to obtaining informed consent and providing a detailed explanation of the study procedures, we implemented software features to prevent the overuse of the EMS system. We designed the games to limit the duration of EMS actuation. We also included kill switches in the system, giving participants a sense of control over the EMS system (like the voice control), even though it controlled their hand movements.
    \item \textbf{Beneficence} is the second aspect that asks designers to question: what benefits are created for participants? The participants in our study appreciated the opportunity to engage with a novel EMS system in a fun and engaging manner, as suggested by our average interview time (section \ref{sec4.3:poststudyphase}). They reflected on the experience of loaning control over their actions to technology and saw value in such experiences. Furthermore, they appreciated a better understanding of their bodies through the fused spectatorship experience.
    \item \textbf{Respect for autonomy} is the third aspect that asks designers to question: how is the participants' autonomy protected? We used the pre-study phase to personalise the experience for each participant. We trimmed the electrodes for every participant so that the EMS could precisely target specific muscles (since each participant's muscle structure is unique) to actuate the gross- and fine-motor movements necessary for spectating the games. Our study also respected participants' autonomy by allowing them to withdraw from the study at any point.
\end{itemize}

\subsubsection{Ethical Design Considerations to Mitigate Potential Harm when Users Loan their Bodily Autonomy with a Computer}\label{sec7.4:Discussion - DesignEthics}
In this section, we draw on literature from HCI and game design to offer suggestions for ethically designing systems that take away player autonomy and provide possible solutions to prevent potential harm.

\begin{itemize}
    \item \textbf{Agency and the balance between control and engagement:} As our games directly influence the participant's bodily movements, they inherently impact their agency \cite{Kaptelinin_Nardi_2012}. Designers could consider ways to balance control and engagement in the system design. One way to mitigate any potential harm that might be caused by the computer to which they are loaning bodily control is to offer a way for a more ``adjustable'' autonomy \cite{Mostafa_Ahmad_Mustapha_2019}. This means that designers could allow players to choose when to activate the EMS or determine the level of influence it has on their movements. In our system, the participant was given a choice to decide when to start the EMS and to pause it at any time. We also used a commercial EMS device, allowing them to control its intensity, pulse rate and width easily. This way, the player maintains a sense of control while still experiencing the unique aspects of the fused spectatorship experience.
    \item \textbf{Privacy concerns:} Physiological actuation of the body, such as involuntary muscle movements in our case, raises concerns about privacy \cite{Lopes_Chuang_Maes_2021}. Designers could be transparent about what physiological data is being collected and used, giving them a sense of ownership and control over their data \cite{Cavoukian_2010}. Moreover, incorporating privacy by design principles can help ensure the system respects user privacy and minimises potential data breaches or misuse risks \cite{Cavoukian_2010}.
    \item \textbf{The risk of overuse:} The interactive nature of our games can lead users into a state of flow \cite{Csikszentmihalyi_Csikzentmihaly_1990}, potentially leading to overuse of our system. This is especially not advisable with EMS as it is recommended to not use it longer than one hour per day, splitting it into two 30-minute sessions \cite{Knibbe_Alsmith_Hornbak_2018}. To mitigate the risk of overusage, designers could incorporate features that encourage responsible use, such as setting time limits, offering reminders to take breaks, or providing self-monitoring tools for tracking usage patterns \cite{Monge_Roffarello_De_Russis_2019}. Our software application notified participants when it detected 30-minutes of continuous EMS use.
\end{itemize}

\section{LIMITATIONS AND FUTURE WORK}\label{sec8:Limitations}
We acknowledge that our work has limitations, specifically concerning EMS calibration, which depends on one's body composition (as highlighted by prior work \cite{Knibbe_Alsmith_Hornbak_2018,Lopes_Ion_Mueller_Hoffmann_Jonell_Baudisch_2015}), which led to some participants being unable to have the full experience of spectating all games. We also acknowledge the sample size (n=12), which speaks to similar sizes in related qualitative studies around games and play \cite{Andres_de_Hoog_Mueller_2018,Li_Patibanda_Brandmueller_Wang_Berean_Greuter_Mueller_2018,Patibanda_Mueller_Leskovsek_Duckworth_2017,La_Delfa_Baytas_Patibanda_2020}. While there were only 12 participants, Fig. \ref{fig:demographics} shows that they played each game a significant number of times. The minimum number of times participants played each game was seven (which was the requirement), and the maximum was 15. Hence, we believe that our work still holds value and suggest it be used to structure future more extensive studies. We also acknowledge that only three out of twelve participants had no prior EMS experience in our study. Therefore, future work could also explore conducting studies with more participants without prior EMS experience to check if the spectator experiences are altered in any way.

Regarding negotiating agency, we suggest that future research can focus on the design of EMS controls in the games. In the presented games, we offered the participants different, but standardised, ways to control the EMS system (Section \ref{sec7.4:Discussion - DesignEthics}). Nonetheless, to suit the different \textit{Fused Spectatorship} types and the associated sense of agency, it might be worthwhile to offer EMS controls based on the person's bodily awareness and ability to loan bodily control to the system. In this way, we could further ensure that the participant is comfortable and confident in sharing bodily control with the system. Lastly, while ``The Entertainer'' experimented with ways to spectate the single-player games with other people, future work could look at deliberately designing social fused spectatorship experiences and performances, for example, by adding affective loops \cite{Robinson_Wiley_Rezaeivahdati_Klarkowski_Mandryk_2020}.

\section{CONCLUSION}\label{sec9:Conclusion}
In conclusion, \textit{Fused Spectatorship} is a novel approach where spectators loan bodily control to a computational Electrical Muscle Stimulation (EMS) system to more actively engage their body in spectating gameplay, beyond the vicarious nature of traditional spectatorship. By designing and evaluating three games with 12 participants, we could articulate four spectator experience themes and four fused spectator types. Our study revealed that participants could not distinguish if they were watching or playing. As such, our work challenges the traditional definition of play that sees ``control'' as essential for facilitating play. Instead, our work suggests that it might be more important that the player's body is engaged than that the player is in control for play to emerge. We also discussed the ethical design considerations associated with our approach. Our findings potentially inform future research and development of fused spectatorship experiences also for other areas besides entertainment, such as medical rehabilitation, and inclusive play for people with impairments. Taken together, we believe that our findings highlight an exciting possibility for designing future fused spectatorship experiences.

\begin{acks}
Rakesh Patibanda, Aryan Saini, Nathalie Overdevest, Elise van den Hoven and Florian `Floyd' Mueller thank the Australian Research Council (Discovery Project Grant - DP190102068). Florian `Floyd' Mueller also thanks the Australian Research Council for DP200102612 and LP210200656. We would also like to thank the participants who volunteered for our study. Additionally, we thank those participants who kindly agreed to be featured in the photographic material included in this publication. Explicit consent was obtained to use their images.
\end{acks}

\bibliographystyle{ACM-Reference-Format}
\bibliography{fused_spec}



\end{document}